\documentclass[pra,twocolumn,preprintnumbers,amsmath,amssymb]{revtex4}

\usepackage{graphicx}
\usepackage{epstopdf}
\usepackage{amscd}

\usepackage{dcolumn}
\usepackage{bm}

\newcommand{\proj}{\operatorname{proj}}  

\newcommand{\vect}[1] {\ensuremath{\mathbf{#1}}} 

\newcommand{\mment}[1] {\ensuremath{\mathbf{#1}}} 

\newcommand{\cvect}[1] {\textsf{#1}} 

\newcommand{\cmatrix}[1]{\textsf{#1}} 

\newcommand{\rmatrix}[1]{\ensuremath{#1}} 

\newcommand{\ev}[1]{\ensuremath{\langle #1 \rangle}} 
\newcommand{\evb}[1]{\ensuremath{\bigl\langle #1 \bigr\rangle}} 

\newcommand{\comm}[2]{\ensuremath{[\cmatrix{#1}, \cmatrix{#2}]}} 

\addtocounter{secnumdepth}{1} 

\begin{document}


\title{Origin of the Correspondence Rules of Quantum Theory}

\author{Philip Goyal}
    \email{pgoyal@perimeterinstitute.ca}
    \affiliation{Perimeter Institute \\ Waterloo, Canada}


\begin{abstract}

To apply the abstract quantum formalism to a particular physical system, one must specify the precise form of the relevant measurement and symmetry transformation operators.   These operators are determined by a set of rules, the \emph{correspondence rules} of quantum theory.  The physical origin of these rules is obscure, and their physical interpretation and their degree of generality is presently unclear.  In this paper, we show that all of the commonly-used correspondence rules can be systematically derived from a new physical principle, the~\emph{Average-Value Correspondence Principle}.     This principle shows that the correspondence rules result from the systematic translation of relations between measurement results known to hold in a classical model of a system, providing these rules with a clear physical interpretation, and clearly demarcating their domain of applicability.

\end{abstract}

\maketitle

\section{Introduction}

In order to apply the standard von Neumann--Dirac quantum formalism to a particular physical system such a particle, one must specify the precise form of the operators that represent measurements performed on the system~(such as position and momentum measurements) or that represent particular symmetry transformations~(such as displacement or rotation) of the frame of reference in which the system is being observed.   These operators are usually determined by reference to a set of rules, the \emph{correspondence rules} of quantum theory.  However, the physical origin of many of these rules is obscure.  Accordingly, the physical meaning and the generality of even of the simplest of these rules is often unclear.  

For example, the conventional operator rules of quantum theory assert that, if measurements~$A$ and~$B$ are represented by operators~$\cmatrix{A}$ and~$\cmatrix{B}$, respectively, then a measurement of~$A+B$ is represented by the operator~$\cmatrix{A}+\cmatrix{B}$~(the \emph{operator sum rule}). Although this seems entirely reasonable on a formal, symbolic level, it is not clear in what \emph{physical} sense~$\cmatrix{A}+\cmatrix{B}$ can be said to `represent' the measurement.  To see the difficultly, consider that a classical measurement of~$S_x + S_z$ is typically implemented by an arrangement where measurements of~$S_x$ and~$S_z$ are performed immediately after one another, and their values added.  When modeled in the quantum framework, the possible outputs of this arrangement, in the case where the measurements are performed on a spin-1/2 particle, are~$+\hbar, 0$ and~$-\hbar$.  However, the operator~$\cmatrix{S}_x + \cmatrix{S}_z$, which is supposed to represent a measurement of~$S_x + S_z$, has only two possible outcomes~$\pm \hbar/\sqrt{2}$.

Perhaps more seriously, the conventional operator rules can lead to inconsistencies.  For example, when the rule that the operator representing a measurement of~$AB$ is given by~$(\cmatrix{A}\cmatrix{B} + \cmatrix{B}\cmatrix{A})/2$ is applied to obtain the operator representing a measurement of~$A^2B$, one can obtain two operators,~$(\cmatrix{A}^2\cmatrix{B} + 2\cmatrix{A}\cmatrix{B}\cmatrix{A} + \cmatrix{B}\cmatrix{A}^2)/4$ and~$(\cmatrix{A}^2\cmatrix{B} + \cmatrix{B}\cmatrix{A}^2)/2$,
 which are not, in general, equivalent~\footnote{See, for example,~\cite{vNeumann55}~(Sec.~IV.1), \cite{Bohm51}~(Sec.~9.12---9.15), and~\cite{Isham-QT}~(Sec.~5.2.1).  We shall discuss one such example in Sec.~\ref{sec:inconsistencies}.}.  

Similar difficulties exist with the canonical commutation relationships.  For example, the relation~$[\cmatrix{x},\cmatrix{p}_x]=i\hbar$ is typically justified by reference to the abstract arguments given by Dirac~\cite{Dirac58}, or to arguments that depend on the  symmetry group of flat spacetime~\cite{TFJordan75,TFJordan69, Ballentine98}.   The former arguments have been found to be logically problematic as well as being physically obscure~\cite{Dickson-Dirac-talk}, and the latter arguments assume the operator rules and, furthermore, do not hold in curved spacetime, leaving it unclear whether the commutation relationships continue to hold there also.    
Similarly, the commutation relation~$[\cmatrix{L}_x, \cmatrix{L}_y]=i\hbar \cmatrix{L}_z$ is ordinarily derived in the infinite-dimensional quantum formalism for a particle~(by transposing the classical relation~$\cmatrix{L}_z = xp_y - yp_x$, and cyclic permutations thereof, into the quantum framework using the operator rules), and is then assumed, without further justification, to also hold for finite-dimensional systems with intrinsic angular momentum.

The elucidation of the physical basis of the correspondence rules of quantum theory is important in order to ensure that the formalism is correctly applied in physical situations other than in which the quantum formalism was generated.  For example, it is important to know whether the position--momentum commutation relations are valid in general spacetimes.  However, the question of what physical ideas underpin the correspondence rules has received relatively little attention, with most recent work on the elucidation of the physical origin of the quantum formalism~(see~\cite{Goyal-QT2c} for references) tending to focus exclusively on the the abstract quantum framework, or being concerned with the derivation of the Schroedinger or Dirac equation directly~\cite{Frieden-Schroedinger-derivation,MJWHall-Schroedinger-derivation,Reginatto-Schroedinger-derivation}. 

In this paper, we show that all of the commonly-used correspondence rules of quantum theory can be obtained systematically from a physical principle which we call the \emph{Average-Value Correspondence Principle}~(AVCP), first briefly introduced in~\cite{Goyal-QT2c}.    In particular, we show that it is possible to derive all of the following types of rules:
\begin{itemize}
\item[(i)]\emph{Operator Rules.} Rules for writing down operators
representing measurements of observables that, in the framework of
classical physics, are known functions of other, elementary,
observables whose operators are given~%
\footnote{The basic operator rules of quantum theory
are~(i)~Rule~1~(Function rule):~If~$A \rightarrow\cmatrix{A}$, then~$f(A)\rightarrow f(\cmatrix{A})$; (ii)~Rule~2~(Sum rule):~If~$A \rightarrow\cmatrix{A}$ and~$B\rightarrow\cmatrix{B}$, then~$f_1(A) + f_2(B) \rightarrow f_1(\cmatrix{A})
+ f_2(\cmatrix{B})$~(and similarly for more than two observables),
and (iii)~Rule~3~(Product rule):~If~$A \rightarrow\cmatrix{A}$ and~$B\rightarrow\cmatrix{B}$,
where~$[\cmatrix{A}, \cmatrix{B}]=0$, then~$f_1(A)f_2(B) \rightarrow f_1(\cmatrix{A})f_2(\cmatrix{B})$~(and similarly for more than two observables).  A more general rule that is often employed
is:~(iv)~Rule~3$'$~(Hermitization rule):~If~$A \rightarrow\cmatrix{A}$ and~$B\rightarrow\cmatrix{B}$,
then~$f_1(A)f_2(B) \rightarrow (f_1(\cmatrix{A})f_2(\cmatrix{B}) +
f_2(\cmatrix{B})f_1(\cmatrix{A}))/2$.}.
For example, such rules are needed to be able to write down the
operator that represents a measurement of~$H$, given the classical
relation~$H = p_x^2/2m + V(x)$, in terms of the
operators~$\cmatrix{x}$ and~$\cmatrix{p}_x$ which represent
measurements of~$x$ and~$p_x$, respectively.

\item[(ii)]\emph{Measurement Commutation Relations.}  Commutation relations
between operators that represent measurements of fundamental
observables such as position, momentum, and components of angular
momentum.  The commutation
relations~$[\cmatrix{x},\cmatrix{p}_x]=i\hbar$ and~$[\cmatrix{L}_x,
\cmatrix{L}_y]=i\hbar\cmatrix{L}_z$, are the obvious examples, while
Dirac's Poisson bracket rule,~$[\cmatrix{A}, \cmatrix{B}] = i\hbar
\widehat{\{A, B\}}$, is the more general rule, where~$\{A,B\}$ is the classical Poisson
bracket for observables~$A$ and~$B$,
and~$\widehat{\{A,B\}}$, $\cmatrix{A}$, and~$\cmatrix{B}$ are the
respective operators.

\item[(iii)]\emph{Measurement--Transformation Commutation Relations.} The
commutation relations between measurement operators and transformation
operators. For example, in the case of the $x$-displacement
operator,~$\cmatrix{D}_x$, one has, for example,~$[\cmatrix{x},\cmatrix{D}_x]=i$ and~$[\cmatrix{p}_x, \cmatrix{D}_x]=0$.  

\item[(iv)]\emph{Transformation Operators.}  Explicit forms of
the operators that represent symmetry transformations of a frame
of reference, such as the $x$-displacement operator~$\cmatrix{D}_x
= -i\, d/dx$.

\end{itemize}

The Average-Value Correspondence Principle is based on the simple idea that the classical and quantum models of any physical system must agree on average.  To illustrate the key idea behind the principle, consider the classical and quantum models of a particle.  Suppose that, in the classical model of a particle,  a measurement of the position,~$x'$, of a particle of mass~$m$, is made at time~$t+\delta t$.  This measurement can be implemented by performing measurements of its position,~$x$, and its momentum,~$p_x$, at time~$t$, and then computing~$ x' = x + p_x \delta t/m$.    Now consider the quantum model of the particle.  Consider three distinct experiments on the particle.  In the first, a measurement of position,~$x$, is performed at time~$t$; in the second, a measurement of momentum,$p_x$, is performed at time~$t$; in the third, a measurement of position,~$x'$, is performed at time~$t$.   The AVCP asserts that, over many runs of these three quantum experiments, the \emph{average} values, which we shall respectively denote~$\ev{x}, \ev{p}_x$ and~$\ev{x'}$, satisfy the relation~$\ev{x'} = \ev{x} + \ev{p}_x \delta t/m$.  That is, \emph{on average}, the classical relation~$ x' = x + p_x \delta t/m$ is satisfied.

This remainder of this paper is organized as follows. We begin in Sec.~\ref{sec:AVCP} by describing the~AVCP. In Sec.~\ref{sec:deduction}, the AVCP is used to obtain several \emph{generalized operator rules} which connect the
average values of operators at different times, from which the commonly used operator rules of quantum theory follow as a special case. 

Next, in Sec.~\ref{sec:commutation-relations}, we use the AVCP to derive many of the commonly employed correspondence rules of quantum theory, namely~(a) the commutation relations~$[\cmatrix{x},\cmatrix{p}_x]=i\hbar$ and~$[\cmatrix{L}_x, \cmatrix{L}_y]=i\hbar\cmatrix{L}_z$, and Dirac's Poisson bracket rule, (b)~the explicit form of the operators for displacements and rotations, and (c)~the commutation relations between momentum and displacement operators, and between angular momentum and rotation operators.   In Sec.~\ref{sec:induction}, we describe a line of argument which naturally leads to the AVCP, and which suggests how it may be generalized.  The paper concludes with a discussion of the results obtained.

We note that the treatment of the correspondence rules is illustrative rather than exhaustive, so that only the most commonly encountered measurement and transformation operators which are needed to formulate non-relativistic and relativistic quantum mechanics have been discussed. Many other correspondence rules~(such as the operators that represent Galilei transformations, temporal displacement, and discrete transformations such as spatial inversion) can be obtained by arguments which closely follow those presented. 

\section{The Average-Value Correspondence Principle}
\label{sec:AVCP}

Suppose that, in a classical experiment, a measurement of observable~$A$ on a physical system can be implemented by an arrangement where measurements of  observables~$A', A'',\dots$ are performed on the system~(possibly performed at a different time to the measurement of~$A$), such that the value obtained from measurement of~$A$ can be calculated as a function,~$f$, of the values obtained from the measurements of~$A', A'',\dots$.   The AVCP asserts that we can then construct a corresponding quantum experiment, where quantum measurements~$\mment{A}, \mment{A}', \mment{A}'',\dots$, corresponding to classical measurements of~$A, A', A'',\dots$, are performed on copies of the same physical system, and asserts that the same functional relation holds \emph{on average} between the values obtained from the quantum measurements, where the average is taken over infinitely many trials of the quantum experiment.  This connection holds only if~$f$ satisfies a particular condition.  The particular form of the quantum experiment is stipulated by the AVCP, and consists in specifying whether, for each pair of quantum measurements, the measurements in the pair are performed on the same or different copies of the system.

The precise form of the corresponding quantum experiment is arrived at by carefully considering a sequence of particular examples, for which the reader is referred to Sec.~\ref{sec:induction}.   Through such considerations, one finds that the key difficulty is what the quantum experiment must be in the case where the operators involved are non-commuting.  For example, suppose that, in the classical framework, a measurement of~$C$ at time~$t$ can be implemented by performing measurements of~$A$ and~$B$ at time~$t$.  Suppose further that, in the quantum framework, it is known that operators~$\cmatrix{A}$ and~$\cmatrix{B}$ represent measurements of~$A$ and~$B$ respectively, and that~$[\cmatrix{A}, \cmatrix{B}]\neq 0$.  In this case,  the AVCP requires that the quantum measurements are performed on separate copies of the system.  On the other hand, if the operators commute, and if the quantum measurements are not performed on sub-systems of a composite system, then the measurements are required to be performed on the same copy of the system.   The intuition here is simple:~when the operators commute, this is an essentially classical situation, and the measurements can be performed on the same copy of the system; when, however, the operators do not commute, this is a distinctively non-classical situation, in which case, to avoid the complication of one measurement's results being affected by that of the other, and to avoid the issue of measurement ordering, the measurements are performed on separate copies of the system.

\subsection*{Statement of the Average-Value Correspondence Principle}
\label{eqn:AVCP-statement}

\begin{figure}[!h]
\begin{centering}
\includegraphics[width=3.25in]{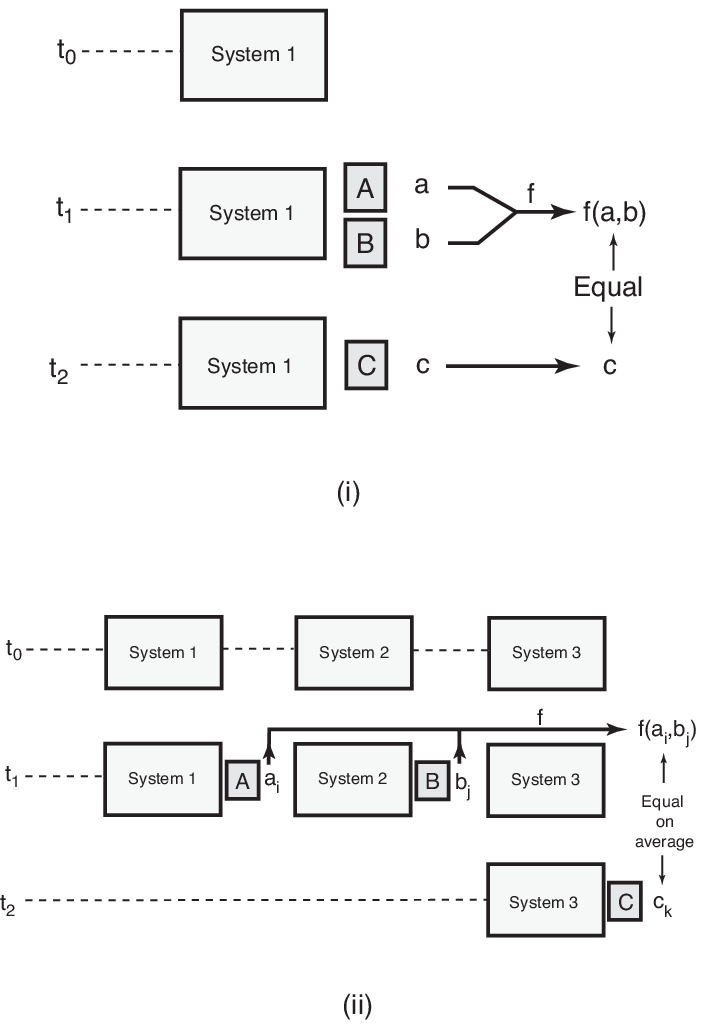}
\caption{\label{fig:AVCP-example} An example of the application of
the AVCP.  (i) A classical experiment showing the measurements
of~$A, B$ and~$C$ performed at times~$t_A, t_B$, and~$t_C$,
respectively, with values denoted as~$a, b$, and~$c$,
respectively. Here,~$t_A = t_B = t_1$ and~$t_C = t_2$. Suppose that
one finds that the relation~$c=f(a,b)$ holds for all initial states
of the system. (ii) The corresponding quantum experiment. Three
copies of the system are prepared in the same initial
state,~$\cvect{v}_0$, at time~$t_0$, and are placed in identical
backgrounds.  In this example, it is assumed that the
operators~$\cmatrix{A}$ and~$\cmatrix{B}$ do not commute. Hence, by
the AVCP, measurements~\mment{A} and~\mment{B} are performed on
different copies of the system. Measurement~\mment{C} is performed
on a separate copy of the system.  In any given run of the
experiment, the probabilities that measurements~\mment{A}, \mment{B}
and~\mment{C} yield values~$a_i, b_j$ and~$c_k$~($i, j, k =
1, \dots, N$), are~$p_i, p'_j$ and~$p_k''$, respectively. The AVCP
then asserts that, provided the polynomial expansion of~$f(a,b)$
contains no product terms involving~$a$ and~$b$, the
relation~$\overline{c}=\overline{f(a,b)}$ holds for all initial
states,~$\cvect{v}_0$, of the system, where the average is taken
over an infinite number of runs of the experiment; that is, $\sum_k
c_k p_k'' = \sum_{ij} f(a_i,b_j) p_i p'_j$ for all~$\cvect{v}_0$.}
\end{centering}
\end{figure}

Consider a classical idealized experiment in which a system~(possibly a composite system) is prepared in some state at time~$t_0$, and is allowed to evolve in a given background . Suppose that a measurement of~$A^{(m)}$~$(m\ge 2)$, performed on the system at time~$t_2$ with value~$a^{(m)}$, can be implemented by an arrangement where measurements of~$A^{(1)}, A^{(2)}, \dots, A^{(m-1)}$ are performed upon one copy of the system at time~$t_1$, and the values of their respective results,  denoted~$a^{(1)}, a^{(2)}, \dots, a^{(m-1)}$, are then used to compute the output~$f(a^{(1)}, a^{(2)}, \dots, a^{(m-1)})$, where~$f$ is an analytic function, so that the relation
\begin{equation*}
\label{eqn:avc}
    a^{(m)}=f(a^{(1)}, a^{(2)}, \dots, a^{(m-1)}) \tag{$*$}
\end{equation*}
holds for all initial (classical)~states of the system~(see Fig.~\ref{fig:AVCP-example}).

Consider the case where the quantum measurements~$\mment{A}^{(1)}, \mment{A}^{(2)}, \dots, \mment{A}^{(m)}$, with operators~$\cmatrix{A}^{(1)}, \cmatrix{A}^{(2)}, \dots, \cmatrix{A}^{(m)}$, represent the measurements of~$A^{(1)}, A^{(2)}, \dots, A^{(m)}$, respectively.  Then, consider the following idealized quantum experimental arrangement consisting of several set-ups, each consisting of identical sources and backgrounds, where, in each set-up, a copy of the system is prepared in the same initial state,~$\cvect{v}_0$, at time~$t_0$.

In one set-up, only measurement~$\mment{A}^{(m)}$ is performed~(at time~$t_2$) and, for any~$i,j$ with~$i\neq j$ and~$i,j \leq m-1$, the measurements~$\mment{A}^{(i)}, \mment{A}^{(j)}$ are performed~(at time~$t_1$) in:
	\begin{enumerate}
		\item the \emph{same} set-up if~$\bigl[\cmatrix{A}^{(i)},  \cmatrix{A}^{(j)}\bigr]=0$, and
    	\item \emph{different} set-ups if~$\bigl[\cmatrix{A}^{(i)},  \cmatrix{A}^{(j)}\bigr]\ne 0$.
	\end{enumerate}
	
Let the values of the results of the measurements~$\mment{A}^{(1)}, \dots, \mment{A}^{(m)}$ in any given run of the experimental arrangement be denoted~$a^{(1)}, \dots, a^{(m)}$, respectively.  The function~$f(a^{(1)}, a^{(2)}, \dots, a^{(m-1)})$ is defined as \emph{simple} provided that its polynomial expansion contains no terms involving a product of eigenvalues belonging to measurements whose operators do not commute.  If~$f$ is simple, then~\eqref{eqn:avc} holds on average, the average being taken over an infinitely large number of runs of the experiment.

\section{Generalized Operator Rules}
\label{sec:deduction}

We will now apply the AVCP to derive operator relations which hold
when the function~$f$ takes various useful forms.  In each
instance of~$f$, we shall first derive a generalized operator rule
which relates the expected values of the relevant operators at
\emph{different times}.  Then, taking the special case when the
expected values are computed at the same time, we obtain the
corresponding operator rule which relates the operators
themselves.

We shall consider a classical experiment where a system is subject
to measurements of~$A$ and~$B$ at time~$t_1$, and to a measurement
of~$C$ at time~$t_2$. We shall suppose that a measurement of~$C$, with
value~$c$, can be implemented by an arrangement in which the
measurements of~$A$ and~$B$ are performed, with respective
values~$a$ and~$b$, and the function~$f(a,b)$ then computed, so
that the relation
\begin{equation}
c= f(a,b)
\end{equation}
holds for all initial states of the system.

In the quantum counterpart of the appropriate experimental arrangement, let
the operators that represent the corresponding quantum measurements be
denoted~$\cmatrix{A}$,~$\cmatrix{B}$, and~$\cmatrix{C}$,
respectively.  To simplify the presentation, we shall only consider
the case where these operators have finite dimension,~$N$; the
results obtained below can be readily shown to apply in the infinite
dimensional case. Let the elements of orthonormal sets of
eigenvectors of~$\cmatrix{A}, \cmatrix{B}$ and~$\cmatrix{C}$ be
denoted~$\cvect{v}_i, \cvect{v}_j'$, and~$\cvect{v}''_k$,
respectively~$(i, j, k = 1,2, \dots, N)$, let the corresponding
eigenvalues be denoted~$a_i, b_j$ and~$c_k$, and let the
probabilities of the~$i$th,~$j$th and~$k$th results of
measurements~$\mment{A}$,~$\mment{B}$ and~$\mment{C}$ in any given
experimental arrangement be denoted by~$p_i$,~$p'_j$ and~$p_k''$,
respectively.

\subsection*{Case 1. $f$ is a function of~$a$ only.}
\label{sec:AVCP-case1}

In this case, the quantum experiment simply consists of two set-ups, involving two copies of the system,
where~$\mment{A}$ is performed on one copy at time~$t_1$ and~$\mment{C}$ on the other copy at time~$t_2$. Since function~$f$ is simple, by the AVCP, the relation
\begin{equation} \label{eqn:AVCP-case-1-main-eqn}
    \sum_k c_k p_k'' =  \sum_i f(a_i) p_i
\end{equation}
holds for all initial states,~$\cvect{v}_0$, of the system.  Explicitly,
\begin{equation}
\begin{aligned}
p_i &= \big|\cvect{v}_i^\dagger \cvect{v}_{t_1}\big|^2 \\
p_k'' &= \big|\cvect{v}_k''^\dagger \cvect{v}_{t_2}\big|^2,
\end{aligned}
\end{equation}
with~$\cvect{v}_t$ being the state of the relevant copy of the system at time~$t$.  Hence, we can write
Eq.~\eqref{eqn:AVCP-case-1-main-eqn} as
\begin{equation}
\cvect{v}^\dagger_{t_2}
                        \bigg(
                        \sum_k \cvect{v}_k'' {\cvect{v}''_k}^\dagger
                        c_k
                        \bigg)
\cvect{v}_{t_2} = \cvect{v}^\dagger_{t_1} \bigg(
                        \sum_i \cvect{v}_i \cvect{v}_i^\dagger f(a_i)
                        \bigg)
\cvect{v}_{t_1}.
\end{equation}
Noting that
\begin{equation}
\begin{aligned}
f(\cmatrix{A}) &= \sum_i \cvect{v}_i \cvect{v}_i^\dagger f(a_i) \\
\cmatrix{C} &= \sum_k \cvect{v}_k'' {\cvect{v}''_k}^\dagger c_k,
\end{aligned}
\end{equation}
we obtain the relation
\begin{equation} \label{eqn:AVCP-case-1-main-result}
\ev{\cmatrix{C}}_{t_2} = \evb{f(\cmatrix{A})}_{t_1},
\end{equation}
which holds for all~$\cvect{v}_0$.  We can summarize the above result in the form of the \emph{generalized function rule}:
\begin{equation} \label{eqn:AVCP-case-1-general-rule}
c(t_2) = f\left(a(t_1)\right) \quad\mapsto\quad \ev{\cmatrix{C}}_{t_2} =
\evb{f(\cmatrix{A})}_{t_1} \quad\forall \cvect{v}_0,
\end{equation}
where, for clarity, the times at which the results are obtained has been explicitly indicated.
In the special case where~$t=t_1 = t_2$, we obtain the usual operator rule, the \emph{function rule}:
\begin{equation} \label{eqn:AVCP-case-1-special-case-result}
c = f(a) \quad\mapsto\quad  \cvect{C} =  f(\cvect{A}).
\end{equation}

\subsection*{Case 2. $f(a,b)= f_1(a) +  f_2(b)$}
\label{sec:AVCP-case2}

It is necessary to consider two sub-cases.  First, if measurements~\mment{A} and~\mment{B} have commuting operators, then, by the AVCP, they are performed on the same copy of the system in the quantum experiment.  Since~$f$ is simple, the AVCP applies, so that
\begin{equation}
\begin{split}
    \sum_k c_k p_k''    &=  \sum_i \bigg( f_1(a_i)  + \sum_j  f_2(b_j)
                                                             p'_{j|i}
                                    \bigg) p_i \\
                        &=   \sum_i \big( f_1(a_i)  +  f_2(b _i)
                                    \big) p_i,
\end{split}
\end{equation}
holds for all initial states,~$\cvect{v}_0$, of the system.  Here, the notation~$p'_{j|i}$ is the probability that
measurement~\mment{B} yields result~$j$ given that~\mment{A} has yielded result~$i$; in this case,~$p'_{j|i} = \delta_{ij}$.  From the above relation, we obtain the generalized operator relation
\begin{equation} \label{eqn:AVCP-case2-av-relation}
\ev{\cmatrix{C}}_{t_2} = \evb{f_1(\cmatrix{A})}_{t_1} +
\evb{f_2(\cmatrix{B})}_{t_1},
\end{equation}
which holds for all initial states,~$\cvect{v}_0$.  In the special case where~$t_1=t_2$, we obtain the operator relation,
\begin{equation} \label{eqn:C-operator1}
    \cmatrix{C} =  f_1(\cmatrix{A}) +  f_2(\cmatrix{B}).
\end{equation}

Second, if measurements~\mment{A} and~\mment{B} have non-commuting operators, then, by the AVCP, they are performed on different copies of the system in the quantum experiment. Since~$f$ is simple, the AVCP again applies, so that the relation
\begin{equation}
    \sum_k c_k p_k''    = \sum_i   f_1(a_i) p_i + \sum_j f_2(b_j) p_j'
\end{equation}
holds for all initial states of the system, which yields the same relation as in Eq.~\eqref{eqn:AVCP-case2-av-relation}.

Hence, combining the foregoing sub-cases, we obtain the \emph{generalized sum rule}:
\begin{multline} \label{eqn:AVCP-case-2-general-rule}
c(t_2) = f_1\left(a(t_1)\right) + f_2\left(b(t_1)\right)  \\
\mapsto \ev{\cmatrix{C}}_{t_2} = \evb{f_1(\cmatrix{A})}_{t_1} +  \evb{f_2(\cmatrix{B})}_{t_1}.
\end{multline}
In the special case where~$t=t_1 = t_2$, we obtain the \emph{sum rule}:
\begin{equation} \label{eqn:AVCP-case-2-special-case-result}
c = f_1(a) + f_2(b) \quad\mapsto\quad  \cvect{C} =  f_1(\cvect{A}) +  f_2(\cvect{B}).
\end{equation}
More generally, in the case of a classical experiment where measurements of~$A^{(1)}, A^{(2)}, \dots, A^{(m-1)}$ are performed on a system at time~$t_1$ and a measurement of~$A^{(m)}$ at time~$t_2$, with
values~$a^{(1)}, \dots, a^{(m)}$, respectively, the AVCP implies the generalized operator rule:
\begin{equation} \label{eqn:AVCP-case2-av-relation-general-av}
a^{(m)}(t_2) = \sum_{l=1}^{m-1} f_l\bigl(a^{(l)}(t_1)\bigr)
\mapsto \evb{\cmatrix{A}^{(m)}}_{t_2} = \sum_{l=1}^{m-1} \evb{f_l(\cmatrix{A}^{(l)})}_{t_1}.
\end{equation}
Taking the special case of simultaneous measurements~($t_1 = t_2$), we obtain the operator rule
\begin{equation} \label{eqn:AVCP-case2-av-relation-general}
a^{(m)} = \sum_{l=1}^{m-1} f_l(a^{(l)}) \quad\mapsto\quad \cmatrix{A}^{(m)} = \sum_{l=1}^{m-1} f_l(\cmatrix{A}^{(l)}).
\end{equation}

\subsection*{Case 3. $f(a,b)= f_1(a)f_2(b)$}
\label{sec:AVCP-case3}

We again consider two sub-cases. First, if measurements~\mment{A} and~\mment{B} are represented by commuting operators, then, in the quantum experiment, they are performed on the same copy of the system. Since~$f$ is, therefore, simple, the AVCP applies, so that the relation
\begin{equation}
\begin{split}
    \sum_k c_k p_k''    &=  \sum_{i,j}  f_1(a_i)f_2(b_j)  p_i
                                                            p'_{j|i} \\
                        &=   \sum_i f_1(a_i) f_2(b_i) p_i
\end{split}
\end{equation}
holds for all initial states,~$\cvect{v}_0$, of the system.  Hence, the generalized operator relation
\begin{equation}
\ev{\cmatrix{C}}_{t_2}  =
\evb{f_1(\cmatrix{A})f_2(\cmatrix{B})}_{t_1}
\end{equation}
holds for all~$\cvect{v}_0$.  In the case where measurements~\mment{A}, \mment{B}, and~\mment{C} are simultaneous, we obtain the operator relation
\begin{equation}
    \cmatrix{C} = f_1(\cmatrix{A}) f_2(\cmatrix{B}).
\end{equation}

Second, if measurements~\mment{A} and~\mment{B} are represented by non-commuting operators, then the function~$f$ is not simple, and the AVCP does not apply.

We can combine the foregoing sub-cases to obtain the \emph{generalized product rule}:
\begin{multline} \label{eqn:AVCP-case-3-general-rule}
c(t_2) = f_1\bigl(a(t_1)\bigr)f_2\bigl(b(t_1)\bigr)    \\
\mapsto
\ev{\cmatrix{C}}_{t_2} = \evb{f_1(\cmatrix{A})
                                f_2(\cmatrix{B})}_{t_1} \quad \text{if}~[\cmatrix{A},
                                \cmatrix{B}]=0.
\end{multline}
In the special case where~$t_1 = t_2$, we obtain the \emph{product rule}:
\begin{equation} \label{eqn:AVCP-case-3-special-case-result}
c = f_1(a)f_2(b) \quad\mapsto\quad  \cmatrix{C} =  f_1(\cmatrix{A})  f_2(\cmatrix{B})
                                             												   \quad \text{if}~[\cmatrix{A}, \cmatrix{B}]=0.
\end{equation}

\subsection*{Some comments on inconsistencies}
\label{sec:inconsistencies}

As mentioned in Sec.~\ref{sec:induction}, if the AVCP is weakened to allow  a measurement of~$AB$ to be represented by an operator in the case where~$[\cmatrix{A}, \cmatrix{B}]\ne 0$, one is lead to a rule that is often stated, namely
\begin{equation} \label{eqn:hermitization-rule}
f_1(a)f_2(b) \mapsto \frac{1}{2} \bigl(f_1(\cmatrix{A}) f_2(\cmatrix{B}) + f_1(\cmatrix{A})f_2(\cmatrix{B}) \bigr)
\end{equation}
However, using this rule, one finds that inconsistencies quickly arise.  For example, one can first apply this rule to find that the operator representing a measurement of~$AB$ is
\begin{equation} \label{eqn:hermitized-form-AB2}
\widehat{AB} = (\cmatrix{A}\cmatrix{B} + \cmatrix{B}\cmatrix{A})/2,
\end{equation}
where the notation~$\widehat{X}$ is used to denote the operator that represents a measurement of~$X$. One can then apply the rule a second time to find the operator that represents a measurement of~$A^2B$. By treating this measurement as a measurement of~$A(AB)$, or as a measurement of~$(A^2)B$, one obtains, respectively, either
\begin{equation}
\begin{split}
\widehat{A(AB)} &=\frac{1}{2} (\cmatrix{A}\widehat{AB} +
                                \widehat{AB}\cmatrix{A}) \\
                &=\frac{1}{4}
                \left(\cmatrix{A}(\cmatrix{A}\cmatrix{B} +
                \cmatrix{B}\cmatrix{A}) +
                (\cmatrix{A}\cmatrix{B} +
                \cmatrix{B}\cmatrix{A})\cmatrix{A} \right) \\
                &=\frac{1}{4} \left(\cmatrix{A}^2\cmatrix{B} +
                2\cmatrix{A}\cmatrix{B}\cmatrix{A} +
                \cmatrix{B}\cmatrix{A}^2 \right),
\end{split}
\end{equation}
or
\begin{equation}
\begin{split}
\widehat{(A^2)B} &=\frac{1}{2} (\widehat{A^2}\cmatrix{B} +
                                \cmatrix{B}\widehat{A^2}) \\
                &= \frac{1}{2} (\cmatrix{A}^2\cmatrix{B} +
                                \cmatrix{B}\cmatrix{A}^2),
\end{split}
\end{equation}
which are, in general, inequivalent.  Hence, the AVCP cannot be applied to non-simple functions of observables, even in weakened form, without leading to inconsistencies.

We also remark that, given the AVCP, it cannot consistently be maintained that every classical measurement is represented by a quantum measurement.  To see this, suppose that every classical measurement is represented by a quantum measurement.  Under this supposition, the function and sum rules can be applied to a measurement of~$(A+B)^2$, with~$[\cmatrix{A}, \cmatrix{B}]\ne 0$, to derive Eq.~\eqref{eqn:hermitized-form-AB2} as follows. First, defining~$d= a+b$, we use the sum rule to find~$\cmatrix{D} = \cmatrix{A} + \cmatrix{B}$, and then use the function rule to find that
\begin{equation}
\begin{split}
\widehat{D^2} &= \cmatrix{D}^2 \\
                &= \cmatrix{A}^2 + \cmatrix{A}\cmatrix{B} +
                    \cmatrix{B}\cmatrix{A} + \cmatrix{B}^2.
\end{split}
\end{equation}
Second, since, by assumption, every classical measurement is represented by a quantum measurement, it follows that, in particular, a measurement of~$AB$ is represented by a quantum measurement.  Therefore, we can use the sum rule directly to find a measurement of~$D^2 = A^2 + 2AB + B^2$:
\begin{equation} \label{eqn:D-squared-2}
\begin{split}
\widehat{D^2} &= \widehat{A^2} + 2\widehat{AB} + \widehat{B^2} \\
                &= \cmatrix{A}^2 + 2\widehat{AB} + \cmatrix{B}^2.
\end{split}
\end{equation}
Equating these expressions for~$\widehat{D^2}$, we obtain Eq.~\eqref{eqn:hermitized-form-AB2}, which, as we have seen, leads to an inconsistency.  Hence, if the AVCP is accepted as valid, the original supposition must be false.  On the assumption that the AVCP is valid, this inconsistency can only be avoided if we conclude that a measurement of~$AB$ cannot be represented by a quantum measurement when~$[\cmatrix{A}, \cmatrix{B}]\ne 0$, in which case the sum rule cannot be applied to obtain Eq.~\eqref{eqn:D-squared-2}.

In summary, given that measurements of~$A$ and~$B$ are represented by quantum measurements~$\mment{A}$ and~$\mment{B}$, one can use the AVCP to find quantum measurements that represent measurements of~$f(A), f_1(A) + f_2(B)$ and, for~$[\cmatrix{A}, \cmatrix{B}]=0$, of~$f_1(A)f_2(B)$; and, more generally, one can find quantum measurements that represent measurements of~$f(A,B)$
when~$f$ is simple.  However, the AVCP also implies that a measurement of~$AB$ cannot be represented by a quantum measurement if~$[\cmatrix{A}, \cmatrix{B}]\ne 0$.

\section{Representation of Measurements and Symmetry Transformations}
\label{sec:commutation-relations}

In this section, we shall use the AVCP to obtain~(a)~the commutation relations~$[\cmatrix{x},\cmatrix{p}_x]=i\hbar$ and~$[\cmatrix{L}_x, \cmatrix{L}_y]=i\hbar\cmatrix{L}_z$~(and cyclic permutations
thereof), and a restricted form of Dirac's Poisson bracket rule, (b)~the explicit form of the displacement and rotation operators, and (c)~the relation between the momentum and displacement operators, and between the angular momentum and rotation operators.

\subsection{Position, momentum, and displacement operators}
\label{sec:position-momentum-relations}

We shall proceed in five steps.  First, by considering a particular system~(a particle moving along the $x$-axis), we obtain the commutation relationship,~$[\cmatrix{x}, \cmatrix{p}_x]=i \hbar$. Second, we obtain the explicit co-ordinate representation of the displacement operator~$\cmatrix{D}_x = \frac{1}{i}
\frac{d}{dx}$.  Third, we obtain the commutation relations~$[\cmatrix{x}, \cmatrix{D}_x]=i$ and~$[\cmatrix{p}_x, \cmatrix{D}_x]=0$.  Fourth, we derive the relationship~$\cmatrix{D}_x = \cmatrix{p}_x/ \hbar$, and thereby obtain the co-ordinate representation of~$\cmatrix{p}_x$. Fifth, we show that the relations obtained are generally valid.

\subsubsection{The position--momentum commutation relationships}
\label{sec:position-momentum-relations-1}

Consider a massless particle moving in the $+x$-direction, where measurements of the $x$-component of position, the $x$-component of momentum, and the energy, are represented by the operators~$\cmatrix{x},
\cmatrix{p}_x, \cmatrix{H}$, respectively.

First, to determine the relationship of~$\cmatrix{H}$ to the operators~$\cmatrix{x}$ and~$\cmatrix{p}_x$, we make use of the fact that the relation~$H = cp_x$, where~$c$ is the speed of light, holds for all classical states~$(x,p_x)$ of the system, so that, from the function rule, it follows that~$\cvect{H} = c\cvect{p}_x$.

Next, to obtain a relation between~$\cmatrix{x}$ and~$\cmatrix{p}_x$, we make use of the fact that, in the quantum model, the expected value of~$x$ at time~$t+\delta t$ can be calculated in two separate ways. First, from the definition of~$\ev{x}_t$, the relation
\begin{equation}
\begin{split}
 \ev{\cmatrix{x}}_{t+\delta t} &= \bigl\langle \cmatrix{U}_t^\dag(\delta t) \,\cmatrix{x}\, \cmatrix{U}_t(\delta t) \bigr\rangle_t       \\
                    &= \left\langle \left( 1+ \frac{i\cmatrix{H}\delta t}{ \hbar}\right)
                    \cmatrix{x}
                        \left( 1- \frac{i\cmatrix{H}\delta t}{ \hbar}\right) \right\rangle_t  + O(\delta t^2) \\
                    &= \ev{\cmatrix{x}}_t + \frac{i}{ \hbar} \delta t \ev{\cmatrix{Hx} - \cmatrix{xH}}_t      + O(\delta t^2)     \\
                    &= \ev{\cmatrix{x}}_t - \frac{ic}{  \hbar} \delta t \evb{[\cmatrix{x},\cmatrix{p}_x]}_t + O(\delta
                    t^2)
\end{split}
\end{equation}
holds for all states,~$\cvect{v}$, of the system, where~$\cmatrix{U}_t(\delta t) = \exp\left(-i\cmatrix{H} \delta t/\hbar\right)$ is the unitary matrix that represents temporal evolution of the system during~$[t, t+\delta
t]$.  Second, using the generalized function rule, it follows from the classical relation~$x(t+\delta t) = x(t) + c\delta t + O(\delta t^2)$ that the relation
\begin{equation}
\ev{\cmatrix{x}}_{t+\delta t} = \ev{\cmatrix{x}}_t + c\delta t +
O(\delta t^2)
\end{equation}
holds for all~$\cvect{v}$.

Equating the above two expressions for~$\ev{x}_{t+\delta t}$, we obtain
\begin{equation}
    \cvect{v}^\dagger \left[\cvect{x}, \cvect{p}_x\right] \cvect{v} = i \hbar
\end{equation}
for all~$\cvect{v}$, which implies that
\begin{equation} \label{eqn:x-p-commutation-relation}
    \left[\cvect{x}, \cvect{p}_x \right] = i \hbar.
\end{equation}
We note that, if one instead considers a particle of mass~$m$, moving non-relativistically in the~$x$ direction, so that the classical Hamiltonian given by~$H=p_x^2/2m + V(x)$, and~$x(t+\delta t) = x(t) + p_x(t)\delta t/m$, then the above computation yields the commutation relation~$\cmatrix{p}_x \left[\cmatrix{x}, \cmatrix{p}_x \right] +   \left[\cmatrix{x}, \cmatrix{p}_x \right]  \cmatrix{p}_x= 2i \hbar\cmatrix{p}_x$, which can be solved to yield Eq.~\eqref{eqn:x-p-commutation-relation}.

Although a particular system has been used to obtain this commutation relation, we shall later present an argument for its general validity.

\subsubsection{Co-ordinate representation of the displacement operator.}

Suppose that, in frame~$S$, the system is in state~$\psi(x)$.  The probability density function over~$x'$ in the frame~$S'$, which is displaced a distance~$-\epsilon$ along the $x$-axis, can be calculated in two equivalent ways, according to whether the transformation from frame~$S$ to~$S'$ is treated as a passive or
active transformation.  Accordingly, the probability density function over~$x'$ can be obtained by performing measurements of~$x'$ in frame~$S'$ upon the system in state~$\psi(x)$, or by performing measurements of~$x$ in frame~$S$ upon the system in the transformed state,~$\exp(-i\epsilon\cmatrix{D}_x)\psi(x)$, and
substituting~$x$ for~$x'$ in the resulting probability density function over~$x$.

First, in frame~$S'$, let us calculate the probability density function over~$x'$ directly.  In this frame, the
operator~$\cmatrix{x}'$ represents a measurement of~$x'$.  In the classical model of the system, the relation
\begin{equation}
x' = x + \epsilon
\end{equation}
holds for all states~$(x,p_x)$ of the system.  Hence, by the function rule, we obtain the operator relation
\begin{equation}
\cmatrix{x}' = \cmatrix{x} + \epsilon.
\end{equation}
Hence, an eigenstate of~$\cmatrix{x}$ with eigenvalue~$x$ is an eigenstate of~$\cmatrix{x}'$ with eigenvalue~$x' = x + \epsilon$.  Therefore, if a measurement of~$x$ on a system in state~$\psi(x)$ yields values in the interval~$[x, x+ \Delta x]$ with probability~$|\psi(x)|^2\Delta x$, then a measurement of~$x'$ on a system in the same state yields values in the interval~$[x', x' + \Delta x']$ with probability density
\begin{equation} \label{eqn:path1}
\Pr\left(x'|S', \psi(x) \right) =  \left|\psi(x'-\epsilon)\right|^2.
\end{equation}

Second, in frame~$S$, measurement~$x$ is performed on the system in the transformed state~$\exp(-i\epsilon\cmatrix{D}_x) \psi(x)$, so that the probability density function over~$x$ is
\begin{equation} \label{eqn:path2}
\Pr\left(x|S, \exp(-i\epsilon\cmatrix{D}_x)\psi(x) \right) =  \left|\exp(-i\epsilon\cmatrix{D}_x) \psi(x)\right|^2.
\end{equation}

The probability density functions over~$x'$ in Eq.~\eqref{eqn:path1} and over~$x$ in Eq.~\eqref{eqn:path2}, must agree under the correspondence~$x \leftrightarrow x'$.  Hence
\begin{equation}
\psi(x-\epsilon) = e^{i\phi(x)} \exp(-i\epsilon\cmatrix{D}_x)
                \psi(x),
\end{equation}
with~$\phi(x)$ being an arbitrary real-valued function of~$x$, which is satisfied for any~$\epsilon$ if and only if~$\phi(x) = 0$ and
\begin{equation} \label{eqn:displacement-operator}
\cmatrix{D}_x = \frac{1}{i} \frac{d}{dx}.
\end{equation}

\subsubsection{The position-displacement and momentum-displacement commutation relations.}

In a classical model, the state of a particle subject to measurements of $x$-position and the $x$-component of momentum, is given by~$(x_0, p_{x0})$ in some frame of reference,~$S$. Consider the following two experiments.

In the first experiment, measurements of the $x$~components of position and momentum of the particle are made in a reference frame,~$S'$, that is displaced by a distance~$\epsilon$ along the $-x$~axis, resulting in the state~$(x',p_x')$ of the particle relative to the co-ordinates of frame~$S'$. According to the classical model,
\begin{equation} \label{eqn:classical-displacement-relations}
    \begin{aligned}
    x' &= x_0 + \epsilon  \\
    p_x' &= p_{x0}.
    \end{aligned}
\end{equation}
In the second experiment, the particle is displaced a distance~$\epsilon$ in the $+x$-direction, and measurements of position and momentum are then performed in frame~$S$, giving the state,~$(x, p_x)$, of the particle in frame~$S$ as~$(x_0 + \epsilon, p_{x0})$.

In classical physics, for all states of the particle, the state~$(x', p_x')$, determined by measurements in frame~$S'$ upon the undisplaced particle, is numerically identical to the state~$(x, p_x)$, determined by measurements in frame~$S$ upon the displaced particle.  That is,
\begin{equation} \label{eqn:classical-active-passive-equivalence}
    (x', p_x') = (x, p_x)
\end{equation}
for all states,~$(x_0, p_{x0})$, of the particle.

Now consider a quantum model of the particle subject to measurements of~$x$ and~$p_x$, and let the state of the particle be given by~$\cvect{v}_0$ in frame~$S$. Consider the first experiment. From Eqs.~\eqref{eqn:classical-displacement-relations}, it follows from the generalized function rule that, in the quantum model of the particle, the relations
\begin{equation} \label{eqn:displacment-av-relations1}
    \begin{aligned}
    \ev{\cmatrix{x}'} &= \cvect{v}_0^\dagger  \cvect{x} \cvect{v}_0 + \epsilon \\
    \ev{\cmatrix{p}_x'} &= \cvect{v}_0^\dagger \cvect{p}_x \cvect{v}_0,
    \end{aligned}
\end{equation}
hold for all quantum states,~$\cvect{v}_0$, of the system.

In the second experiment, the displacement of the particle is a continuous, symmetry transformation of the system, and therefore can be represented by a unitary transformation of the state,~$\cvect{v}_0$, and, in particular, by the operator~$\exp\left( -i\epsilon \cvect{D}_x \right)$, where~$\cvect{D}_x$ is a Hermitian operator.  To first order in~$\epsilon$, measurements of~$x$ and~$p_x$ performed on this state have expected values
\begin{equation} \label{eqn:displacment-av-relations2}
    \begin{aligned}
    \ev{\cmatrix{x}} &= \cvect{v}_0^\dagger (1+i\epsilon \cvect{D}_x) \cvect{x}
        																							(1-i\epsilon \cvect{D}_x) \cvect{v}_0 \\
    \ev{\cmatrix{p}_x} &= \cvect{v}_0^\dagger (1+i\epsilon \cvect{D}_x) \cvect{p}_x
        																							(1-i\epsilon \cvect{D}_x) \cvect{v}_0.
    \end{aligned}
\end{equation}

From Eq.~\eqref{eqn:classical-active-passive-equivalence}, by the generalized function rule, the average values~$\ev{x'}$ and~$\ev{p_x'}$ of Eqs.~\eqref{eqn:displacment-av-relations1} are respectively equal to the average values~$\ev{x}$ and~$\ev{p_x}$ of Eqs.~\eqref{eqn:displacment-av-relations2} for all~$\cvect{v}_0$.   Hence, we obtain that the relations
\begin{equation} \label{eqn:displacment-av-relations3}
    \begin{aligned}
    \cvect{v}_0^\dagger (1+i\epsilon \cvect{D}_x) \cvect{x}
        (1-i\epsilon \cvect{D}_x) \cvect{v}_0  &= \cvect{v}_0^\dagger \cvect{x} \cvect{v}_0 + \epsilon \\
    \cvect{v}_0^\dagger (1+i\epsilon \cvect{D}_x) \cvect{p}_x
        (1-i\epsilon \cvect{D}_x) \cvect{v}_0 &= \cvect{v}_0^\dagger  \cvect{p}_x \cvect{v}_0,
    \end{aligned}
\end{equation}
hold for all~$\cvect{v}_0$ to first order in~$\epsilon$, which yield the commutation relations
\begin{equation} \label{eqn:displacment-commutation-relations}
    \begin{aligned}
    \left[ \cvect{x}, \cvect{D}_x \right] &= i \\
    \left[ \cvect{p}_x, \cvect{D}_x \right] &= 0
    \end{aligned}
\end{equation}

\subsubsection{The displacement-momentum relation and the co-ordinate representation of~$\cmatrix{p}_x$.}

From Eqs.~\eqref{eqn:x-p-commutation-relation} and~\eqref{eqn:displacment-commutation-relations}, it follows that
\begin{equation}
\label{sec:displacement-minus-momentum-commutators}
    \begin{aligned}
    \left[\cvect{x}, \left(\cvect{D}_x - \cvect{p}_x/ \hbar \right)\right] &= 0 \\
    \left[\cvect{D}_x, \left(\cvect{D}_x - \cvect{p}_x/ \hbar \right) \right] &= 0.
    \end{aligned}
\end{equation}

Now, in the co-ordinate representation, the operators~$\cmatrix{x}$ and~$\cmatrix{D}_x$ are given by~$x$ and~$-i \hbar\, d/dx$, respectively, and one can readily show that~$\{x, -i \hbar\, d/dx\}$ forms an irreducible set~\footnote{See~\cite{Ballentine98}, Appendix~2.}. By Schur's lemma~\footnote{See, for example, Ref.~\cite{Ballentine98}, Appendix~1.}, it therefore follows from Eqs.~\eqref{sec:displacement-minus-momentum-commutators} that
\begin{equation} \label{eqn:displacement-momentum-relation1}
\cvect{D}_x = \frac{\cvect{p}_x}{ \hbar} + \gamma \cvect{I},
\end{equation}
where~$\gamma$ is real since the operator~$\left(\cvect{D}_x - \cvect{p}_x/ \hbar \right)$ is Hermitian.  For a given displacement,~$\epsilon$, the constant~$\gamma$ results in the same overall shift of phase of any state,~$\cvect{v}$, of a system, and therefore produces no physically observable effects on the system.  Hence,~$\gamma$ can be set equal to zero without any loss of generality, so that we obtain
\begin{equation} \label{eqn:displacement-momentum-relation2}
\cvect{D}_x = \frac{\cvect{p}_x}{ \hbar}.
\end{equation}
Analogous relationships for the displacement operator corresponding to displacements in the~$y$ and~$z$ directions can be obtained in a similar way.

Finally, from Eqs.~\eqref{eqn:displacement-operator} and~\eqref{eqn:displacement-momentum-relation2}, we find
\begin{equation} \label{eqn:x-momentum-operator}
\cmatrix{p}_x = \frac{ \hbar}{i} \frac{d}{dx}.
\end{equation}

\subsubsection{Generality.} \label{sec:generality}

The representations of $x$- and~$p_x$-measurements have been obtained above by considering, in the first step, a quantum model of a particular physical system, namely a particle moving along the $x$-axis.  In the general case of a particle moving in an arbitrary direction, measurements of~$x, y, z$, and~$p_x, p_y, p_z$ are subsystem measurements, and can therefore be represented in the model of the composite system consisting of a particle, subject to measurements chosen from a measurement set generated by a measurement of~$\vect{r}=(x,y,z)$, by the operators~$x, y, z$ and~$-i \hbar \,\partial/\partial x, -i \hbar \,\partial/\partial y, -i \hbar \,\partial/\partial z$, respectively.

These representations of measurements of position and momentum are also more generally valid for other systems, as we shall explain below.

\medskip
\paragraph{State-determined measurements.}

Suppose that, in the classical framework, a measurement of~$A$ is
performed on a system, and the result is determined by the state of
the system alone.  That is, in particular, the result is
independent of the background of the system or of any
parameters~(such as charge or rest mass) that describe intrinsic
properties of the system.  We shall then say that this measurement
is a \emph{state-determined} measurement.  For example, the result
of a position measurement on a particle is determined by the state
of the particle, and is independent of whether or not the particle
is in an electromagnetic field and is independent of the mass or
charge of the particle.  In general, any measurement of an
observable that is a function only of the degrees of freedom of the
state of the system is a state-determined measurement.  In contrast,
a measurement of the total energy of a system is, in general,
dependent upon not only the state of the system, but also upon the
background of the system, and is therefore not a state-determined
measurement.

Now, consider two quantum models of two different physical
systems, system~1 and system~2, in different backgrounds, and suppose that measurements~$\mment{A}_1$ and~$\mment{A}_2$, respectively, are performed on the systems,
where~$\mment{A}_1$ and~$\mment{A}_2$ represent a measurement
of~$A$ performed on the respective systems. Suppose, further, that
the two models have the same dimension. If the (classical)~measurement of~$A$
is state-determined when performed on both systems~1 and~2, then,
by the AVCP, it follows that~$\ev{\cmatrix{A}_1} =
\ev{\cmatrix{A}_2}$ holds for all states,~$\cvect{v}$, where
operators~$\cmatrix{A}_1, \cmatrix{A}_2$ represent
measurements~$\mment{A}_1, \mment{A}_2$, respectively. It follows
at once that the operators~$\cmatrix{A}_1, \cmatrix{A}_2$ are
identical.

Hence, provided that two systems admit classical models with respect
to a measurement of~$A$ that is state-determined, and admit quantum
models of the same dimension with respect to
measurements~$\mment{A}_1$ and~$\mment{A}_2$, the operators that
represent~$\mment{A}_1$ and~$\mment{A}_2$ in the respective models
must be identical.

Therefore, if state-determined measurements of~$x$ and~$p_x$ are
performed on any system, then, in a quantum model of the system
subject to measurements in the measurement set containing quantum
measurements that represent measurements of~$x$ and~$p_x$, where
these measurements yield a continuum of possible results, their
representations are the same as those obtained above.  Therefore,
the commutation relations involving~$\cmatrix{x}, \cmatrix{p}_x$
and~$\cmatrix{D}_x$ are also generally valid.  Similar conclusions
clearly hold for measurements of~$y, z$ and~$p_y, p_z$.

Therefore, in the case of a particle where the interaction energy in
the Hamiltonian is obtained from a scalar potential that is
dependent on position only, in which case the measurements of
position and momentum are state-determined, the above
representations are valid. Below, we shall consider a physically
important case where the measurement of momentum is not
state-determined.

\medskip
\paragraph{Particle in a magnetic field.}

In the case of a charged particle in a magnetic field background
described in the Hamiltonian framework, the state of the particle
is~$(\vect{x}, \dot{\vect{x}})$, but the generalized co-ordinates
are taken to be~$(\vect{x}, \vect{p})$, where~$\vect{p} =
m\dot{\vect{x}} + e\vect{A}$, where~$m$ and~$e$ are the mass and
charge of the particle, respectively, and~$\vect{A}=(A_x, A_y, A_z)$
is the vector potential.  In this case,~$\vect{p}$ depends both upon
the state of the particle and the state of the background.
Therefore, a measurement of~$\vect{p}$ is not a state-determined
measurement, and the foregoing argument cannot be used to argue that
the operators representing the measurements of the components,~$p_x,
p_y, p_z$, of~$\vect{p}$ are those derived above. Instead, we reason
as follows.

First, for a particle with state~$(\vect{x}, \dot{\vect{x}})$ in a
magnetic field, in the argument of
Sec.~\ref{sec:position-momentum-relations-1}, the commutation
relation for the $x$-component of the motion in
Eq.~\eqref{eqn:x-p-commutation-relation} becomes
\begin{equation} \label{eqn:x-x-dot-commutator}
\bigl[\cmatrix{x}, m\dot{\cmatrix{x}} \bigr] = i \hbar,
\end{equation}
Then, from~$\vect{p} = m\dot{\vect{x}} + e\vect{A}$, the sum rule
gives
\begin{equation}
\cmatrix{p}_x = m \dot{\cmatrix{x}} + eA_x(\cmatrix{x},
\cmatrix{y}, \cmatrix{z}),
\end{equation}
which, together with Eq.~\eqref{eqn:x-x-dot-commutator}, implies
that
\begin{equation} \label{eqn:canonical-commutator-in-A-field}
\left[\cmatrix{x}, \cmatrix{p}_x \right] = i \hbar,
\end{equation}
as before.

Second, we note that, in the classical framework, the
momentum~$\vect{p}$ as defined above is invariant under
displacement of the reference frame.  Therefore,
Eqs.~\eqref{eqn:displacment-commutation-relations} remain
unchanged, and, using
Eq.~\eqref{eqn:canonical-commutator-in-A-field}, we
obtain~$\cmatrix{D}_x = \cmatrix{p}_x/ \hbar$.    Third, and
finally, the argument leading to the co-ordinate representation
of~$\cmatrix{D}_x$ remains unchanged since the argument only
involves measurements of position, which are state-determined
measurements. Therefore, the explicit representation
of~$\cmatrix{p}_x$ remains that given in
Eq.~\eqref{eqn:x-momentum-operator}, and similarly for the~$y$-
and~$z$-components of the motion.

\subsubsection{Remark on applications.}

The correspondence rules derived above allow the quantum theoretic
modeling of a non-relativistic particle in an arbitrary classical
background consisting of gravitational and electromagnetic fields,
which leads to the non-relativistic Schroedinger equation. In the
case of a multi-particle system, the rules~(not discussed here)
for dealing with identical particles are, additionally, required.

In addition, the above rules allow the modeling of a photon without
consideration of polarization degrees of freedom~(leading to a
complex wave equation), a structureless relativistic
particle~(leading to the Klein-Gordon equation), and a relativistic
particle with internal degrees of freedom~(which, with the
appropriate auxiliary assumptions, leads to the Dirac equation).

\subsection{Angular momentum and rotation operators}

We shall proceed in three steps.  First, we shall obtain the
commutation relation~$[\cmatrix{L}_x, \cmatrix{L}_y] =
i\hbar\cmatrix{L}_z$, and cyclic permutations thereof, up to an
additive constant.  Second, we shall obtain the commutator relations
that hold between the rotation operators,~$\cmatrix{R}_x,
\cmatrix{R}_y, \cmatrix{R}_z$, and the angular momentum operators,
and shall then use these relations to determine the value of the
additive constant.  Third, we shall determine the relations that
hold between the rotation and angular momentum operators, and
indicate how the explicit representations of the angular momentum
operators and rotation operators can be determined.

\subsubsection{Components of Angular Momentum.}

Consider an experimental set-up where, in the classical model of the
set-up, measurements are performed upon a classical spin, with
magnetic moment~$\boldsymbol{\mu}$, which determine the values of
the rectilinear components of angular momentum of the system.
Suppose that the measurements of the components of angular momentum
along the $x$-, $y$-, and~$z$-directions, and the measurement of
energy, are represented by the operators~$\cmatrix{L}_x,
\cmatrix{L}_y, \cmatrix{L}_z$, and~$\cmatrix{H}$, respectively.

In particular, consider a set-up where a magnetic
field,~$\vect{B}$, is applied to the spin.  In the classical model
of this set-up, the energy associated with the interaction
is~$-\boldsymbol{\mu}\cdot\vect{B}$. Since~$\boldsymbol{\mu} = q
\vect{L}/2m$, where~$q$ and~$m$ are the charge and mass,
respectively, of the spin, and~$\vect{L}=(L_x, L_y, L_z)$ is its
angular momentum vector, the energy can be written
as~$-(q/2m)\,\vect{B}\cdot\vect{L}$. By the sum
rule~(Sec.~\ref{sec:AVCP-case2}), the quantum mechanical
Hamiltonian is given by
\begin{equation}
\cmatrix{H} = -\frac{q}{2m} (B_x\cmatrix{L}_x + B_y\cmatrix{L}_y +
B_z\cmatrix{L}_z),
\end{equation}
where~$B_x, B_y$ and~$B_z$ are the rectilinear components
of~$\vect{B}$.

The application of a magnetic field to a classical spin causes its
angular momentum vector,~$\vect{L}$, to rotate about the axis
along which the magnetic field is applied by an angle that is
proportional both to~$|\vect{B}|$ and to the duration for which
the field is applied.  Let the rotation matrix corresponding to a
rotation about axis~$a$ be denoted~$\rmatrix{R}_a(\theta)$,
where~$\theta$ is the angle of rotation. From the properties of
rotation matrices, it follows that
\begin{equation} \label{eqn:rotation-matrix-relations}
 \rmatrix{R}_x (\epsilon) \rmatrix{R}_y (\epsilon) -
    \rmatrix{R}_y (\epsilon) \rmatrix{R}_x (\epsilon) =
 \rmatrix{R}_z (\epsilon^2) - \rmatrix{I},
\end{equation}
where~$\epsilon$ is an infinitesimal angle,
and~$\rmatrix{R}_a(\epsilon)$ is an infinitesimal rotation which can
be implemented by application of a magnetic field~$\vect{B}$ along
the axis~\textit{a} for some time~$\delta t$.  Using this
relationship, it is possible to deduce the commutation relations
that hold between the quantum mechanical operators,~$\cmatrix{L}_x,
\cmatrix{L}_y, \cmatrix{L}_z$, in the following way.

The unitary evolution corresponding to the application of a
magnetic field~$\vect{B}$ to a spin for a time~$\delta t$ is
\begin{equation} \label{unitary-for-magnetic-field}
\cmatrix{U}(\delta t) = \exp \left(-\frac{i}{\hbar} \cmatrix{H}
\delta t \right).
\end{equation}
If magnetic fields of equal strength are applied for equal
times,~$\delta t$, along the $x, y,$ and~$z$-axes, respectively,
the corresponding unitary evolution is given to first order
in~$\delta t$, respectively, by
\begin{equation}
    \begin{aligned}
 \cmatrix{U}_1(\delta t) &= 1- \frac{i\epsilon}{\hbar} \cmatrix{L}_x  \\
 \cmatrix{U}_2(\delta t) &= 1- \frac{i\epsilon}{\hbar} \cmatrix{L}_y  \\
 \cmatrix{U}_3(\delta t) &= 1- \frac{i\epsilon}{\hbar} \cmatrix{L}_z.
    \end{aligned}
\end{equation}

Define~$\proj(\cvect{v})$ as the operation upon the quantum
state,~$\cvect{v}$, of a spin which returns a three-dimensional
vector,~$\ev{\vect{\cmatrix{L}}}$, with
components~$\ev{\cmatrix{L}_x}, \ev{\cmatrix{L}_y}$
and~$\ev{\cmatrix{L}_z}$.

If the application of a magnetic field,~$\vect{B} = B_z \vect{k}$,
say, to a classical spin causes a rotation of~$\vect{L}$ by
angle~$\theta$, then, by the generalized operator rule in
Eq.~\eqref{eqn:AVCP-case2-av-relation-general}, in the quantum model
of the spin, the application of the field rotates the
vector~$\ev{\vect{\cmatrix{L}}} = \proj(\cvect{v})$ by the
angle~$\theta$ about the z-axis. From
Eq.~\eqref{eqn:rotation-matrix-relations}, it therefore follows
that, for any~$\cvect{v}$,
\begin{multline}  \label{eqn:U-matrix-relations}
\proj(\cmatrix{U}_1(\epsilon) \cmatrix{U}_2(\epsilon) \cvect{v}) -
    \proj(\cmatrix{U}_2(\epsilon) \cmatrix{U}_1(\epsilon) \cvect{v} ) \\
    = 
    \proj(\cmatrix{U}_3(\epsilon^2) \cvect{v} ) - \proj (\cvect{v}).
\end{multline}
Using the definitions
\begin{align*}
    \cvect{v}_1 &= \cmatrix{U}_3(\epsilon^2) \cvect{v} =  \cvect{v} + \delta  \cvect{v}_1  \\
    \cvect{v}_2 &= \cmatrix{U}_1(\epsilon)\cmatrix{U}_2(\epsilon) \cvect{v}
                                         =  \cvect{v} + \delta  \cvect{v}_2     \\
    \cvect{v}_3 &= \cmatrix{U}_2(\epsilon)\cmatrix{U}_1(\epsilon) \cvect{v}
                                         =  \cvect{v} + \delta  \cvect{v}_3,     \\
\intertext{where}
    \delta  \cvect{v}_1 &= -\frac{i}{\hbar} \epsilon^2 \cmatrix{L}_z \cvect{v}    \\
    \delta  \cvect{v}_2 &=
        -\left[ -\frac{i}{\hbar} \epsilon (\cmatrix{L}_x + \cmatrix{L}_y)
                -\frac{1}{\hbar^2} \epsilon^2 \cmatrix{L}_x \cmatrix{L}_y
        \right] \cvect{v} \\
    \delta  \cvect{v}_3 &=
        -\left[ -\frac{i}{\hbar} \epsilon (\cmatrix{L}_x + \cmatrix{L}_y)
                -\frac{1}{\hbar^2} \epsilon^2 \cmatrix{L}_y \cmatrix{L}_z
        \right] \cvect{v},
\end{align*}
equation~\eqref{eqn:U-matrix-relations} becomes
\begin{equation}  \label{eqn:U-matrix-relations2}
\proj(\cvect{v} + \delta\cvect{v}_2) - \proj(\cvect{v} +
\delta\cvect{v}_3) = \proj(\cvect{v} + \delta\cvect{v}_1) -
\proj(\cvect{v}).
\end{equation}
Equating the $x$-components of this equation, we obtain
\begin{equation}
\cvect{v}^\dagger\cmatrix{L}_x (\delta \cvect{v}_2 - \delta
\cvect{v}_3) + (\delta \cvect{v}_2 - \delta \cvect{v}_3)^\dagger
\cmatrix{L}_x \cmatrix{v} = \cvect{v}^\dagger\cmatrix{L}_x \delta
\cvect{v}_1  + \delta \cvect{v}_1^\dagger \cmatrix{L}_x
\cmatrix{v},
\end{equation}
and, inserting the explicit forms of the~$\delta \cvect{v}_i$, we
obtain the commutation relation
\begin{subequations}
\begin{equation}  \label{eqn:an-operator-commutes-with-x-y-zA}
    \bigl[ \cmatrix{L}_x , \; \cmatrix{L}_z + \frac{i}{\hbar} [\cmatrix{L}_x,
        \cmatrix{L}_y ] \bigr]  =0.
\end{equation}
Equating the $y$- and $z$-components similarly, one obtains the
relations
\begin{align} \label{eqn:an-operator-commutes-with-x-y-z}
    \bigl[ \cmatrix{L}_y , \; \cmatrix{L}_z + \frac{i}{\hbar} [\cmatrix{L}_x,
        \cmatrix{L}_y ] \bigr] &=0 \\
    \bigl[ \cmatrix{L}_z , \; \cmatrix{L}_z + \frac{i}{\hbar} [\cmatrix{L}_x,
        \cmatrix{L}_y ] \bigr] &=0
        \label{eqn:an-operator-commutes-with-x-y-zC}
\end{align}
\end{subequations}
By inspection, the above commutation relations have the solution
\begin{subequations}
\begin{equation} \label{eqn:L-x-L-y-solution}
\bigl[ \cmatrix{L}_x, \cmatrix{L}_y \bigr] = i \hbar \cmatrix{L}_z
+
                                    i\gamma_1\cmatrix{I},
\end{equation}
where~$\gamma_1$ is real constant since the
operators~$i[\cmatrix{L}_x, \cmatrix{L}_y ]$ and~$\cmatrix{L}_z$ are
hermitian. We shall later show that this solution is, in fact, the
most general one.

The discussion leading to this result can be repeated to yield the
relations

\begin{align}
\bigl[ \cmatrix{L}_y, \cmatrix{L}_z \bigr] &= i \hbar \cmatrix{L}_x
                + i\gamma_2\cmatrix{I} \label{eqn:L-y-L-z-solution} \\
\bigl[ \cmatrix{L}_z, \cmatrix{L}_x \bigr] &= i\hbar\cmatrix{L}_y
        + i\gamma_3\cmatrix{I}.\label{eqn:L-z-L-x-solution}
\end{align}
\end{subequations}
In order to determine the values of $\gamma$-factors, we require
the commutation relations between the rotation
operators,~$\cmatrix{R}_x, \cmatrix{R}_y, \cmatrix{R}_z$,
and~$\cmatrix{L}_x, \cmatrix{L}_y, \cmatrix{L}_z$, which we shall
now derive.

\subsubsection{Rotation--angular momentum commutation relations.}

Let an infinitesimal clockwise rotation of a frame of reference by
angle~$\epsilon$ about the~$z$-axis be represented by unitary
transformation~$\exp(-i\epsilon \cmatrix{R}_z)$,
where~$\cmatrix{R}_z$ is Hermitian.

Now consider a set-up where measurements of~$L_x, L_y$, and~$L_z$
are performed on a system in the original and in the transformed
frame of reference.  In the classical model of this situation, the
results of the measurements performed in the original~(unprimed)
and transformed~(primed) frames are, to first order in~$\epsilon$,
related by
\begin{equation} \label{eqn:classical-rotation-connections}
\begin{pmatrix} L_x' \\ L_y' \\ L_z'\end{pmatrix} =
                                \begin{pmatrix}
                                    1 & -\epsilon & 0 \\
                                    \epsilon & 1 & 0 \\
                                    0 &   0 &   1
                                \end{pmatrix}
                                \begin{pmatrix}
                                    L_x \\ L_y \\ L_z
                                \end{pmatrix}.
\end{equation}

By the generalized operator rule in
Eq.~\eqref{eqn:AVCP-case2-av-relation-general}, it follows that, in
the quantum model of the situation, the relation
\begin{equation} \label{eqn:av-rotation-connections}
        \begin{pmatrix} \ev{\cmatrix{L}_x'} \\ \ev{\cmatrix{L}_y'} \\ \ev{\cmatrix{L}_z'}
        \end{pmatrix}
        =
        \begin{pmatrix}
            1 & -\epsilon & 0 \\
            \epsilon & 1 & 0 \\
            0 &   0 &   1
        \end{pmatrix}
        \begin{pmatrix}
        \ev{\cmatrix{L}_x} \\ \ev{\cmatrix{L}_y} \\ \ev{\cmatrix{L}_z}
        \end{pmatrix}
\end{equation}
holds for all states,~$\cvect{v}$, of the system.  Using the
relation
\begin{equation}
\ev{\cmatrix{L}_x'} = \cvect{v}^\dagger (1+ i\epsilon
\cmatrix{R}_z)\cmatrix{L}_x (1-  i\epsilon \cmatrix{R}_z) \cvect{v}
+ O(\epsilon^2),
\end{equation}
we find from Eq.~\eqref{eqn:av-rotation-connections} that
\begin{equation}
\ev{\cmatrix{L}_x} + i\epsilon \ev{[\cmatrix{R}_z, \cmatrix{L}_x]}
=
            \ev{\cmatrix{L}_x} - \epsilon \ev{\cmatrix{L}_y}
\end{equation}
for all~$\cvect{v}$, which implies that
\begin{subequations}
\begin{equation} \label{eqn:R-z-L-x-commutator}
\bigl[\cmatrix{R}_z, \cmatrix{L}_x \bigr] = i\cmatrix{L}_y.
\end{equation}
Proceeding similarly for~$\ev{\cmatrix{L}_y'}$
and~$\ev{\cmatrix{L}_z'}$, we obtain
\begin{align}
\left[\cmatrix{R}_z, \cmatrix{L}_y\right] &= -i\cmatrix{L}_x \label{eqn:R-z-L-y-commutator} \\
\left[\cmatrix{R}_z,\cmatrix{L}_z\right] &= 0.
                                \label{eqn:R-z-L-z-commutator}
\end{align}
\end{subequations}

The commutation relations for~$\cmatrix{R}_x$ and~$\cmatrix{R}_y$
can be obtained by parallel arguments.

\subsubsection{Angular momentum commutation relations.}

Left-multiplying Eq.~\eqref{eqn:L-z-L-x-solution}
by~$\cmatrix{R}_z$, and applying
Eqs.~\eqref{eqn:R-z-L-x-commutator}--\eqref{eqn:R-z-L-z-commutator},
we obtain
\begin{equation}
\bigl[\cmatrix{L}_z, \cmatrix{L}_x \bigr] \cmatrix{R}_z - i
\bigl[\cmatrix{L}_y, \cmatrix{L}_z \bigr] = \left(i\hbar
\cmatrix{L}_y + i\gamma_3 \cmatrix{I}\right)\cmatrix{R}_z + \hbar
\cmatrix{L}_x.
\end{equation}
Using the Eq.~\eqref{eqn:L-z-L-x-solution} right-multiplied
by~$\cmatrix{R}_z$, this implies that
\begin{subequations}
\begin{equation}
\bigl[\cmatrix{L}_y, \cmatrix{L}_z \bigr] = i\hbar \cmatrix{L}_x.
\end{equation}
Parallel arguments applied to Eqs.~\eqref{eqn:L-x-L-y-solution}
and~\eqref{eqn:L-y-L-z-solution} using the commutation relations
between~$\cmatrix{R}_x, \cmatrix{R}_y$ and~$\cmatrix{L}_x,
\cmatrix{L}_y, \cmatrix{L}_z$ yield
\begin{align}
\bigl[ \cmatrix{L}_z, \cmatrix{L}_x \bigr] &= i\hbar\cmatrix{L}_y \\
\bigl[ \cmatrix{L}_x, \cmatrix{L}_y \bigr] &= i \hbar
                                                    \cmatrix{L}_z.
\end{align}
\end{subequations}

\subsubsection{Explicit form of angular momentum operators}

From the classical relation~$L^2 = L_x^2 + L_y^2 + L_z^2$, it
follows from the sum rule that~$\cmatrix{L}^2 =\cmatrix{L}_x^2
+\cmatrix{L}_y^2+\cmatrix{L}_z^2$.  Using this relation and the
above commutation relations for~$\cmatrix{L}_x, \cmatrix{L}_y$
and~$\cmatrix{L}_z$, explicit representations of these operators for
finite~$N$ can be obtained and the irreducibility of the
representations of~$\cmatrix{L}_x, \cmatrix{L}_y, \cmatrix{L}_z$ can
be shown~\footnote{See~\cite{Group-Theory-in-Physics}~(Ch.~10,
Sec.~3), for example.}. Therefore, by Schur's lemma, the solution
given in Eq.~\eqref{eqn:L-x-L-y-solution} is the most general
solution of
Eqs.~\eqref{eqn:an-operator-commutes-with-x-y-zA}--\eqref{eqn:an-operator-commutes-with-x-y-zC},
and similarly for the solutions given in
Eqs.~\eqref{eqn:L-y-L-z-solution} and~\eqref{eqn:L-z-L-x-solution}.

Although the representations of~$\cmatrix{L}_x, \cmatrix{L}_y$
and~$\cmatrix{L}_z$ have been obtained by considering a particular
physical system, they are generally valid on account of the
argument given in Sec.~\ref{sec:generality}. Therefore the
commutation relations for~$\cmatrix{L}_x, \cmatrix{L}_y$
and~$\cmatrix{L}_z$ are generally valid.

\subsubsection{Rotation--angular momentum relations,
and explicit form of the rotation operators.}

Using the commutation relationships for~$\cmatrix{L}_x,
\cmatrix{L}_y$ and~$\cmatrix{L}_z$, it follows from
Eqs.~\eqref{eqn:R-z-L-x-commutator}--\eqref{eqn:R-z-L-z-commutator}
that~$(\hbar \cmatrix{R}_z - \cmatrix{L}_z)$ commutes
with~$\cmatrix{L}_x, \cmatrix{L}_y$, and~$\cmatrix{L}_z$.
Since~$\{\cmatrix{L}_x, \cmatrix{L}_y, \cmatrix{L}_z\}$ is an
irreducible set, it follows from Schur's lemma that
\begin{equation}
\hbar \cmatrix{R}_z - \cmatrix{L}_z = \gamma \cmatrix{I},
\end{equation}
where~$\gamma$ is a real-valued constant since~$\cmatrix{R}_z$
and~$\cmatrix{L}_z$ are Hermitian.  For any given~$\epsilon$, a
non-zero value of~$\gamma$ results in the same overall change of
phase for all states transformed
by~$\exp(-i\epsilon\cmatrix{R}_z)$, and so cannot give rise to
observable consequences. Hence,~$\gamma$ can be set to zero
without loss of generality. Therefore,~$\cmatrix{R}_z =
\cmatrix{L}_z/\hbar$, and, similarly,~$\cmatrix{R}_x =
\cmatrix{L}_x/\hbar$ and~$\cmatrix{R}_y = \cmatrix{L}_y/\hbar$.

Using the explicit representations of~$\cmatrix{L}_x,
\cmatrix{L}_y, \cmatrix{L}_z$ for any given dimension~$N$, the
explicit representation of the rotation operators follows at once
from these rotation--angular momentum relations. The explicit
co-ordinate representations of the rotation operators in the
infinite-dimensional case can also be determined by an argument
similar to that used earlier to determine the explicit form of the
displacement operators.

\subsection{Commutators and Poisson brackets}
\label{sec:PB-analogy}

In this section, we shall obtain a relation between the Poisson
Bracket,~$\{A, B\}$, and the commutator~$[\cmatrix{A},
\cmatrix{B}]$, where~$A$ and~$B$ are the classical observables of
a physical system describable in the classical Hamiltonian
framework, and~$\cmatrix{A}, \cmatrix{B}$ are the operators that
represent measurements of these observables. Dirac's Poisson
Bracket rule asserts the relation
\begin{equation} \label{eqn:Dirac-PB}
[\cmatrix{A}, \cmatrix{B}] = i\hbar \widehat{\{A, B\}},
\end{equation}
where~$\widehat{\{A,B\}}$ is the operator that represents a
measurement of~$\{A, B\}$.  Below, we shall derive this relation
using the AVCP in the case where~$B$ is the Hamiltonian.

Consider the Hamiltonian model of a system with state~$(q_1,
\dots, q_N; p_1, \dots, p_N)$ where~$N\ge 1$. The temporal rate of
change of the function~$F(q_1, \dots, q_N; p_1, \dots, p_N)$ is
given in terms of the Hamiltonian,~$H(q_1, \dots, q_N; p_1, \dots,
p_N)$, by
\begin{equation} \label{eqn:F-dot-PB}
\begin{split}
\dot{F} &=  \{F,H\} \\
        & = \sum_{i=1}^{N} \left\{ \frac{\partial F}{\partial q_i}
            \frac{\partial H}{\partial p_i} -  \frac{\partial H}{\partial q_i}
            \frac{\partial F}{\partial p_i} \right\}.
\end{split}
\end{equation}
Consider the quantum model of the system with state~$\cvect{v}$,
where the measurements of the~$q_i$ and the~$p_i$ are represented by
operators~$\cvect{q}_i$ and~$\cvect{p}_i$, respectively. If~$H$ is
simple, then, by the AVCP, a measurement of~$H$ can be represented
by the operator~$\cmatrix{H}$; otherwise, according to the AVCP, it
is not possible to describe the temporal evolution of the system in
the quantum model. If both of the functions~$\dot{F}$ and~$\{F, H\}$
are simple, then, by the AVCP, they are represented by the
operators~$\widehat{\dot{F}}$ and~$\widehat{\{F,H\}}$, respectively,
and from Eq.~\eqref{eqn:F-dot-PB}, by the generalized function rule,
the relation
\begin{equation} \label{eqn:PB1}
\evb{\widehat{\dot{F}}}_t =  \evb{\widehat{\{F,H\}}}_t,
\end{equation}
holds for all~$\cvect{v}$.

Now, in the classical model, the function~$\dot{F}$ is defined,
for all states, as
\begin{equation} \label{eqn:classical-F-dot}
\dot{F} =  \lim_{\Delta t \rightarrow 0} \left\{ \frac{F(t+\Delta
t) - F(t)}{\Delta t} \right\}.
\end{equation}
If~$F(t)$ and~$F(t+\Delta t)$ are both simple, then, according to
this definition,~$\dot{F}(t)$ is also simple, and, using the
generalized sum rule~(regarding the measurement of~$F(t+\Delta t)$
as the one being implemented in terms of measurements of~$F(t)$
and~$\dot{F}(t)$), we obtain the relation
\begin{equation} \label{eqn:PB2}
\evb{\widehat{\dot{F}}}_t =\lim_{\Delta t \rightarrow 0}
\frac{1}{\Delta t}
                    \big\{ \ev{\cmatrix{F}}_{t+\Delta t} - \ev{\cmatrix{F}}_t
                    \big\},
\end{equation}
which holds for all~$\cvect{v}$, with the operator~$\cmatrix{F}$
representing a measurement of~$F$.

If the functions~$\{F, H \}$ and~$F(t)$ are both simple, then,
since~$F(t+\Delta t) = F(t) + \{F, H \} \Delta t$, it follows
that~$F(t+\Delta t)$ is also simple. In that case, both
Eqs.~\eqref{eqn:PB1} and~\eqref{eqn:PB2} hold for all~$\cvect{v}$.
Equating these two expressions for~$\ev{\widehat{\dot{F}}}_t$, we
obtain the relation
\begin{equation}
\begin{aligned} \label{eqn:PB-average-relation}
    \evb{\widehat{\{F,H\}}}_t   &=\lim_{\Delta t \rightarrow 0} \frac{1}{\Delta t}
                     \bigg\{ \left\langle \left( 1+ \frac{i}{\hbar} \cvect{H}\Delta t\right)
                        \cvect{F}
                     \left( 1- \frac{i}{\hbar} \cvect{H}\Delta t \right)
                     \right\rangle_t    \\
                     &\quad\quad\quad\quad\quad\quad\quad\quad\quad\quad\quad\quad\quad\quad\quad\quad\quad   - \ev{\cmatrix{F}}_t  \bigg\} \\
                &= i\hbar^{-1} \left\langle [\cvect{H},\cvect{F}]
                \right\rangle_t,
\end{aligned}
\end{equation}
which holds for all~$\cvect{v}$, so that
\begin{equation} \label{eqn:PB-analogy}
    i\hbar \widehat{\{F,H\}}  =  [\cvect{F}, \cvect{H}].
\end{equation}
Hence, we obtain Eq.~\eqref{eqn:Dirac-PB} in the special case
where~$\cmatrix{B} = \cmatrix{H}$, subject to the condition that
the functions~$A, B$ and~$\{A,B\}$ are simple.  Using this relationship, we can readily evaluate useful commutation relationships.  For example, setting~$F = x$, and~$H = c p_x$, we find~$\{F,H\} = c$.  Hence, since the functions~$F, H$, and~$\{F,H\}$ are simple, Eq.~\eqref{eqn:PB-analogy} immediately gives~$[\cvect{x}, \cvect{p}_x] = i\hbar$.

If one or more of the functions~$F, H$, and~$\{F,H\}$ is not
simple, then Eq.~\eqref{eqn:PB-analogy} does not follow from the
above argument.
To take a specific example, suppose that, for a system with
state~$(x,p_x)$, where~$[\cmatrix{x}, \cmatrix{p}_x]=i\hbar$, we
choose~$F=x^3$ and~$H=\gamma p_x^3$, where~$\gamma$ is a constant.
We can then apply the function rule to obtain the corresponding
operators~$\cmatrix{F} =\cmatrix{x}^3$ and~$\cmatrix{H} =
\gamma\cmatrix{p}_x^3$,  and use these to find
\begin{equation} \label{eqn:pb-example-1}
[\cmatrix{F}, \cmatrix{H}] = 3i\gamma\hbar \left(\cmatrix{x}^2
\cmatrix{p}_x^2 + \cmatrix{x} \cmatrix{p}_x^2 \cmatrix{x} +
\cmatrix{p}_x^2 \cmatrix{x}^2\right).
\end{equation}
However, the function~$\{F,H\} = 9\gamma x^2p_x^2$ is not simple,
which implies that the AVCP cannot be used to write down an operator
which represents a measurement of~$\{F,H\}$.   If we were
nonetheless to apply the Hermitization rule in
Eq.~\eqref{eqn:hermitization-rule} to a measurement of~$\{F,H\}$~(in
spite of the inconsistencies to which we have shown this would lead)
we would obtain
\begin{equation} \label{eqn:pb-example-2}
i\hbar \widehat{\{F, H\}} = \frac{9i\gamma\hbar}{2}
\left(\cmatrix{x}^2\cmatrix{p}_x^2 + \cmatrix{p}_x^2 \cmatrix{x}^2
\right),
\end{equation}
but this differs from~$[\cmatrix{F}, \cmatrix{H}]$ by the
constant~$2\gamma\hbar^3$.  Since the expected value
of~$\gamma\hbar\cmatrix{x}^2\cmatrix{p}_x^2$ could itself be of
order~$\gamma\hbar^3$, there is no guarantee that the difference
between Eqs.~\eqref{eqn:pb-example-1} and~\eqref{eqn:pb-example-2}
will be negligible.

\section{Origin of the AVCP}
\label{sec:induction}

In this section, we shall summarise the line of argument that leads to the AVCP.

Suppose that there is some measurement apparatus which, as described within the framework of classical physics, performs a measurement of some classical observable~$A$ on a physical system.  And suppose that there is some other measurement apparatus which, as described within the framework of quantum theory, performs quantum measurement~$\mment{A}$ on the same system, and is described by some Hermitian operator,~$\cmatrix{A}$.    Let us suppose further that quantum measurement~$\mment{A}$ is regarded as representing the measurement of the classical observable~$A$.   The question we now wish to start by considering is whether there is a quantum measurement that represents a measurement of~$A^2$, and, if so, what operator represents this measurement.    For simplicity, we suppose that the quantum system has dimension~$N$, and is prepared in a pure state,~$\cvect{v}$.   

\begin{figure}[!h]
\begin{centering}
\includegraphics[width=3.25in]{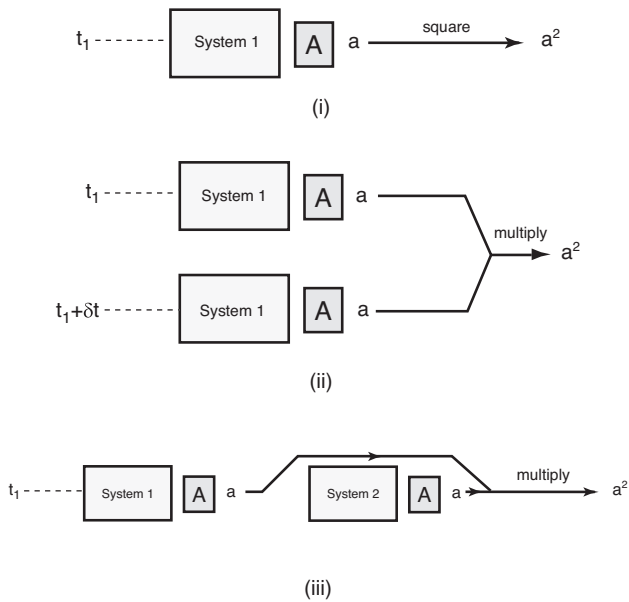}
\caption{\label{fig:mments-of-A-squared} Three arrangement that implement 
a measurement of~$A^2$.
In~(i), a measurement of~$A$ is made on one copy of the system, and
the value obtained is squared to give the output; in (ii), two immediately
successive measurements of~$A$ are made on one copy of the system,
and the output is obtained by multiplying the two values obtained; and
(iii)~two simultaneous measurements of~$A$ are made on two separate
copies of the system prepared in the same state, and the output is
obtained by multiplying the two values obtained. In a classical model of
these arrangements, each arrangement yields the same output. However,
in the quantum counterparts of these arrangements, the expected output
of~(i) and~(ii) is~$\overline{a^2}$ whereas the expected output
of~(iii) is~$(\overline{a})^2$.}
\end{centering}
\end{figure}

To answer this question, we begin by asking how a measurement of~$A^2$ is actually implemented.   Classically, one can imagine a measurement of~$A^2$ being implemented using one of three different arrangements~(see Fig.~\ref{fig:mments-of-A-squared}): (i)~make a measurement of~$A$ on one copy of the system, and square the value obtained; (ii)~make two immediately successive measurements of~$A$ on one copy of the system, and multiply the two values obtained; or~(iii)~make two simultaneous measurements of~$A$ on two separate copies of the system prepared in the same state, and multiply the two values obtained.    All three arrangements yield the same output, and so can be regarded as equally valid implementations of a measurement of~$A^2$.  

Let us now consider the quantum counterparts of these arrangements where a quantum measurement~$\mment{A}$ takes the place of all instances of a measurement of~$A$.  In the first such quantum arrangement, one copy of the system is prepared in some state and measurement~$\mment{A}$ performed on this system, yielding result~$i$~($i=1, 2, \dots, N$), with associated value~$a_i$, with probability~$p_i$.  The output of the arrangement is thus~$a_i^2$ with probability~$p_i$, so that
\begin{equation}
\begin{split}
\label{eqn:expected-outcome-example-1-implementation-1}
\text{Expected output} &= \sum_i (a_i)^2 p_i \\
                                &= \overline{a^2}.
\end{split}
\end{equation}
Similarly, one finds that arrangement~(ii) yields the same expected output. However, in arrangement~(iii), where, in each run of the experiment, two copies of the system are prepared in the same state, one obtains
\begin{equation}
\label{eqn:expected-outcome-example-1-implementation-3}
\begin{split}
\text{Expected output} &= \sum_{i,j} (a_i a_j) p_ip'_j \\
                                &= (\overline{a})^2,
\end{split}
\end{equation}
with~$p_i$ and~$p'_j$ denoting, respectively, the probabilities that measurements~$\mment{A}$ on the first and second copy yield result~$i$ and~$j$~($i, j=1, 2, \dots, N$).

Therefore, although the three arrangements are equivalent in the classical framework, their quantum counterparts are \emph{not}, in general, equivalent in the quantum framework.    This difference arises due to the fact that, in the quantum framework, an immediate repetition of a measurement on the same copy of a system is different from performing a second simultaneous measurement on a separate, identically-prepared copy of the system.

\subsubsection{The average-value condition}
\label{sec:av-condition}

The existence of different arrangements that implement the same classical measurement immediately raises two questions.  First, are the quantum counterparts of these arrangements, in some sense, equally valid in the quantum framework, or it is possible to find some reasonable physical basis upon which to select particular arrangements and regard these as more fundamental than the others? Second, is it possible to find operators that represent the selected arrangements, and, if so, are all of these  represented by the same operator?

To answer these questions, we begin by observing that, although the above arrangements are all regarded as \emph{bone fide} implementations of measurements in the classical model, a measurement is only
describable as such in the quantum framework, and so can be called a quantum measurement, if it can be represented by a Hermitian operator which represents a single measurement performed upon one copy of the system at a particular time. So, for example, although the quantum counterpart of arrangement~(iii)  can be \emph{modeled} in the quantum framework, it \emph{cannot} be described as a quantum measurement since it involves two separate measurements. In contrast, the quantum counterpart of arrangement~(i)  \emph{can} be described as a quantum measurement since it only involves a single measurement on one copy of the system.

However, even though an arrangement involving more than one measurement cannot itself be regarded as a quantum measurement,  we can reasonably ask whether it is possible to find a quantum measurement,~$\mment{C}$, with operator~$\cmatrix{C}$, which, in some sense to be determined, can nonetheless be said to \emph{represent} such an arrangement.

At the outset, we note that it makes no sense to require that measurement~$\mment{C}$ always yield a value that coincides with the output of a given arrangement since measurement results are only probabilistically determined in the quantum framework. However, we \emph{can} impose the simple condition that, over an infinite number of runs,  the average value obtained from measurement~$\mment{C}$ should coincide with the average output obtained from the arrangement for any initial state of the system.

For example, in the case of an arrangement that implements a measurement of~$A^2$, quantum measurement~\mment{C} which, by hypothesis, represents the arrangement, must be such that, for all states of the system,~$\ev{\cmatrix{C}}$ is equal to the expected output obtained from the arrangement.  In the case of arrangements~(i) and~(ii) described above, using Eq.~\eqref{eqn:expected-outcome-example-1-implementation-1}, we accordingly obtain the condition that the relation
\begin{equation}
\begin{split}
\ev{\cmatrix{C}}    &= \overline{a^2} \\
                    &= \ev{\cmatrix{A}^2}
\end{split}
\end{equation}
must hold for all states,~$\cvect{v}$, of the system.  From this condition, we can conclude that
\begin{equation}
\label{eqn:avc-rule-1}
\cmatrix{C} = \cmatrix{A}^2.
\end{equation}
In the case of arrangement~(iii), using Eq.~\eqref{eqn:expected-outcome-example-1-implementation-3}, we obtain the condition that the relation
\begin{equation}
\begin{split}
\ev{\cmatrix{C}}    &= (\overline{a})^2 \\
                    &= \ev{\cmatrix{A}}^2
\end{split}
\end{equation}
must hold for all~$\cvect{v}$.  By diagonalizing~$\cmatrix{A}$, one can readily show that this condition implies that~$\cmatrix{A}$ is a multiple of the identity, which represents a trivial measurement that yields the same result irrespective of the state of the system. Therefore, arrangement~(iii) does not satisfy the above average-value condition in the case of any non-trivial measurement of~$A$, and can therefore be reasonably eliminated. Hence, in this case, the average-value condition is sufficiently strong so as to be able to pick out arrangements~(i) and~(ii).  Furthermore, since the average-value condition also implies that these arrangements are both represented by the operator,~$\cmatrix{A}^2$, one can unambiguously conclude that a measurement of~$A^2$ is represented by the operator~$\cmatrix{A}^2$.

\begin{figure}[!h]
\begin{centering}
\includegraphics[width=3.25in]{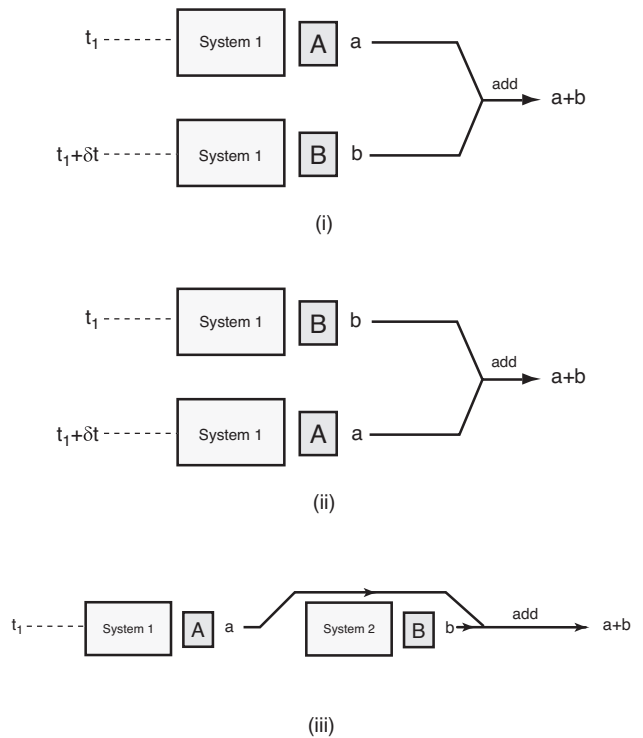}
\caption{\label{fig:mments-of-A-plus-B} Three implementations of a
measurement of~$A+B$.
In~(i), a measurement of~$A$, and then a measurement of~$B$, is made
on one copy of the system, and the values obtained are then added to give
the output; in (ii), a measurement of~$B$, and then a measurement
of~$A$, is made on one copy of the system, and the values obtained are then
added to give the output; and (iii)~simultaneous measurements of~$A$
and~$B$ are made on two separate copies of the system prepared in
the same state, and the output is obtained by adding the two
values obtained. In a classical model of these arrangements, each
arrangement yields the same output. However, the quantum counterparts of these arrangements do not, in general, have the same expected outputs.}
\end{centering}
\end{figure}

We can proceed in a similar way for a measurement of~$A + B$, where measurements of~$A$ and~$B$ are assumed to occur at the same time.  In this case, one needs to take into account the additional fact that, in the quantum framework, the order in which two measurements is performed is of possible importance.  Accordingly, one can imagine implementing a measurement of~$A+B$ using one of at least three arrangements~(see Fig.~\ref{fig:mments-of-A-plus-B}): (i)~make a measurement of~$A$, and then a measurement of~$B$, on one copy of the system, and add the values obtained; (ii) make a measurement of~$B$, and then a measurement of~$A$, on one copy of the system, and add the values obtained; (iii)~make a measurement of~$A$ on one copy of the system and, simultaneously, a measurement of~$B$ on a second copy of the system prepared in the same state as the first copy, and add the values obtained. Once again, in the classical framework, the outputs of these arrangements agree. However, in the quantum counterparts of these arrangements, one finds that, if~$[\cmatrix{A}, \cmatrix{B}]\ne 0$, the expected outputs of all three arrangements will, in general, disagree with one another.

Restricting ourselves for the time being to measurements~$\mment{A}$ and~$\mment{B}$ that are not subsystem measurements~\footnote{A subsystem measurement is defined here to be a measurement performed on a single subsystem of a composite system.}, one finds that, in the case of a measurement of~$C = A+B$, only arrangement~(iii) is possible if~$[\cmatrix{A}, \cmatrix{B}]\ne 0$, which then yields the operator
\begin{equation}
\label{eqn:avc-rule-2} \cmatrix{C} = \cmatrix{A}+\cmatrix{B}.
\end{equation}
If~$[\cmatrix{A}, \cmatrix{B}]= 0$, then all three arrangements are possible, and all yield the same operator,~$\cmatrix{C}$, as above.

Finally, in the case of a measurement of~$AB$, with~$[\cmatrix{A}, \cmatrix{B}]=0$, one finds that the only possible arrangement is the one where measurements~\mment{A} and~\mment{B} are performed on the same copy of the system, in which case the operator~$\cmatrix{A}\cmatrix{B}$ is obtained.

Hence, we see that the average-value condition is sufficient to yield a unique operator representation for the measurements considered above.  Based on the above considerations, we can tentatively formulate the following general rule:~\emph{in the case of a measurement which is implemented by an arrangement that contains two measurements~(not subsystem measurements) that, in its quantum counterpart, are represented by quantum measurements with commuting operators, it is possible to find a quantum measurement that represents the arrangement if the two measurements are performed on the same copy of the system; but, when the operators do not commute, the measurements must be performed on different copies of the system.}

We now consider implementations of a measurement of~$AB$ in the case when~$[\cmatrix{A}, \cmatrix{B}]\ne 0$. The general rule just given suggests that we should consider the arrangement where the measurements of~$A$ and~$B$ are performed on different copies of the system.  The quantum counterpart of this arrangement has the expected output
\begin{equation}
\sum_i \sum_j (a_i b_j) p_i p_j' =
\ev{\cmatrix{A}}\ev{\cmatrix{B}},
\end{equation}
with~$p_i$ and~$p'_j$ denoting respectively the probabilities that the measurement~$\mment{A}$ on the first copy and the measurement~$\mment{B}$ on the second copy yield result~$i$ and~$j$~($i, j=1, 2, \dots, N$), and~$a_i, b_j$ respectively denoting the values of the $i$th and $j$th results of measurements~$\mment{A}$ and~$\mment{B}$.  Imposing the above average-value condition, the operator~$\cmatrix{C}$ that represents this implementation must satisfy the condition
\begin{equation} \label{eqn:measurement-AB-av-desideratum}
\ev{\cmatrix{C}} =  \ev{\cmatrix{A}}\ev{\cmatrix{B}}
\end{equation}
for all~$\cvect{v}$.  However, for non-commuting~$\cmatrix{A}$ and~$\cmatrix{B}$, one finds that~$\cmatrix{C}$ cannot be found such that this relation is satisfied for all~$\cvect{v}$. We note, however, that with
\begin{equation}
\label{eqn:hermitized-form-AB} \cmatrix{C} = \frac{1}{2}
(\cmatrix{A}\cmatrix{B} + \cmatrix{B}\cmatrix{A}),
\end{equation}
equation~\eqref{eqn:measurement-AB-av-desideratum} holds for the eigenstates of~$\cmatrix{A}$ and the eigenstates of~$\cmatrix{B}$, which suggests the possibility of weakening the average-value condition in this case so that we only require that Eq.~\eqref{eqn:measurement-AB-av-desideratum} holds for these eigenstates.  However, as we shall illustrate later~(Sec.~\ref{sec:inconsistencies}), the application of Eq.~\eqref{eqn:hermitized-form-AB} can lead to inconsistencies. Consequently, we conclude that, from the point of view of the average-value condition, it is not possible to find an operator that represents a measurement of~$AB$ when the operators~$\cmatrix{A}$ and~$\cmatrix{B}$ do not commute. More generally, we find that, when~$[\cmatrix{A}, \cmatrix{B}]\ne 0$, it is necessary to exclude measurements of~$f(A,B)$, with~$f$ analytic~(so that~$f$ has a well-defined polynomial expansion in~$A$ and~$B$), where the polynomial expansion of~$f$ contains product terms.

Finally, in the case of a classical measurement which is implemented by an arrangement which contains two measurements that are performed on different subsystems of a composite system, one finds that the quantum counterpart of this arrangement can be represented by a quantum measurement irrespective of whether the two measurements are performed on the same or on different copies of the system, and that the different possible arrangements are represented by the same operator.

\subsubsection{Generalizations}

In the examples above, we have considered arrangements that implement a classical measurement in which the measurement and the measurements in the arrangement are performed at the same time and in the same frame of reference. However, as the following examples illustrate, these are unnecessary restrictions.

First, classically, for a non-relativistic particle of mass~$m$, one can implement a measurement of~$x$ performed at time~$t+\delta t$ by an arrangement where measurements of~$x$ and~$p_x$ are performed at time~$t$, and the function~$x + p_x \,\delta t/m$ is then computed.

Second, one can implement a measurement of~$x'$ on a particle in the reference frame~$S'$ that is displaced along the $x$-axis relative to frame~$S$ by performing a measurement of~$x$ in frame~$S$, and computing the appropriate function~$x'=f(x)$ that relates~$x$ and~$x'$.

The above considerations regarding the implementation of a classical measurement are applicable without change to the case where the measurement and the measurements in the arrangement by which it is implemented are performed at different times or in different frames of reference.  The statement of the AVPC accordingly generalizes our notion of the implementation of a classical measurement.

The AVCP could also be generalized in other ways, for example to the case where the measurements of~$A^{(1)}, \dots, A^{(m-1)}$ are not performed at the same time. However, these generalizations are unnecessary for the derivations of the usual correspondence rules of quantum theory, and are therefore not discussed here.

\section{Discussion}

The derivation given in this paper provides several interesting insights into measurements and their representation.   The first insight is that a measurement of a classical observable~(such as a measurement of~$x^2$) can be implemented by more than one arrangement, and that the quantum counterparts of these arrangements are not, in general, equivalent. 

Second, it is possible to impose a simple average-value condition that must be satisfied by an operator that can be said to represent a quantum arrangement.  This condition implies that many arrangements cannot be represented by an operator, and can therefore be eliminated from consideration. That is, one finds that there are arrangements which, although acceptable representations of a measurement in the classical framework, have quantum counterparts that cannot be represented by operators in the quantum framework without violating a very mild average-value condition.

Third, in the case of a quantum arrangement that satisfies the average-value condition, the operator that represents the arrangement is uniquely determined by the average-value condition provided that the function,~$f$, that describes the arrangement, is simple. One also finds that those arrangements that satisfy the average-value condition are represented by the same operator, so that it is possible to represent a measurement of~$f$ by a unique operator. If~$f$ is not simple, then it does not appear to be possible to apply the average-value condition, even in a weakened form, without inconsistencies arising.

Fourth, we have found that the AVCP is incompatible with the assumption that every classical measurement on a system is represented by a quantum measurement in the quantum model of the system.  For example, the AVCP implies that a measurement of~$AB$ does not have an operator representation if~$\comm{A}{B} \neq 0$.   We leave it as an open question as to whether ways can be found for systematically dealing with the inconsistencies~(as illustrated in Sec.~\ref{sec:deduction}) to which rules such as the Hermitization rule~(which is obtainable by weakening AVCP as indicated in Sec.~\ref{sec:av-condition}) tend to lead.

The fifth insight rests on the fact that, rather surprisingly, the AVCP enables correspondence rules of many different types---operator rules, measurement commutation relations, measurement--transformation commutation relations, and the forms of the transformation operators---to be obtained in a uniform manner. Consequently, one can see that the difference between these types of rules depends simply upon whether the classical relations that one is taking over into the quantum framework are relations between measurements performed at the same time~(leading to the operator rules), at different times~(leading to measurement commutation relations), or in different frames of reference~(leading to measurement--transformation commutation relations and to the explicit forms of transformation operators). In short, from the perspective provided by the derivation, the commutation relation~$[\cmatrix{x}, \cmatrix{p}_x] = i\hbar$ is no more mysterious in its origin than the operator relation~$\cmatrix{H} = \cmatrix{p}_x^2/2m$.

Sixth, the derivation provides a clearer physical foundation to many particular correspondence rules that are commonly used in quantum theory. For example, the commutation relation~$[\cmatrix{L}_x, \cmatrix{L}_y]=i\hbar \cmatrix{L}_z$ is here obtained directly for finite- and infinite-dimensional quantum systems, and in a manner that makes clear its connection with the properties of rotations. Similarly, a restricted form of Dirac's Poisson bracket rule has been derived in a systematic manner using the AVCP without making use of abstract analogies.

More generally, the derivation provided here shows that average-value correspondence is rather useful.  In particular it allows familiar relations known to hold in classical physics~(the less precise theory) to be systematically transformed into exact relations amongst operators~(such as commutation relations) in quantum theory~(the more precise theory), which is something that one might intuitively not expect would be possible.  Moreover, insofar as its use as a methodological tool in building up new physical theories is concerned, the only formal aspect of quantum theory of which the AVCP makes use is the idea of the commutativity or otherwise of two measurement operators, which however can be replaced by a simple operational criterion within the probabilistic operational framework described in~\cite{Goyal-QT2c}.  Hence, the AVCP itself can be regarded as standing quite independently of the quantum formalism, and can be used as a constructive tool in building up new theories within a general operational probabilistic framework.

Finally, we remark that, although the general notion of average-value
correspondence is already familiar in elementary quantum mechanics
through Ehrenfest's theorem, the possibility
that such a correspondence might serve as the basis for a
constructive principle that allows the correspondence rules of
quantum theory to be determined by appropriately-chosen classical
relations does not appear to have been widely
explored~\footnote{%
Refs.~\cite{vNeumann55, Groenewold-Principles-QM, Bohm51} mention
the general idea of average-value correspondence in their
discussion of the operator rules of quantum theory.  For example,
Groenewold~\cite{Groenewold-Principles-QM}~(Eqs.~(1.32)--(1.34))
remarks that the sum rule is equivalent to a condition on the
expectations of the respective operators, but the idea is not
formulated in a manner that is sufficiently systematic to derive
the operator rules, and no attempt is made to derive the any of
the other types of correspondence rule~(such as the measurement commutation relations)
using average-value correspondence.  Bohm~\cite{Bohm51} clearly
articulates the idea that average-value correspondence can be used
as a constraint on quantum theory, and uses it to determine
particular instances of the function rule~(Secs.~9.5--9.21) and to
determine the Hamiltonian operator that represents a
non-relativistic particle~(Secs.~9.24--9.26). However, the idea is
not
systematically formulated and applied beyond these special cases.}.  %
It has been shown here that it is possible to formulate the notion
of average-value correspondence in the form of an exact physical
principle which, roughly speaking, allows the logic of Ehrenfest's
argument to be reversed.

\begin{acknowledgments}

I am indebted to Steve Gull and Mike Payne for their support, and to Tetsuo Amaya, Leslie Ballentine, Matthew Donald, and Gordon Fleming and Yiton Fu for discussions and invaluable comments.  I would like to thank the Cavendish Laboratory; Trinity College, Cambridge; Wolfson College, Cambridge; and Perimeter Institute for institutional and financial support.    Research at Perimeter Institute is supported in part by the Government of Canada through NSERC and by the Province of Ontario through MEDT.

\end{acknowledgments}


\begin{thebibliography}{10}

\bibitem{Dirac58}
P.A.M. Dirac.
\newblock {\em Principles of Quantum Mechanics}.
\newblock Oxford Science Publications, fourth edition, 1999.

\bibitem{TFJordan75}
T.~F. Jordan.
\newblock Why {$-i\nabla$} is the momentum.
\newblock {\em Am. J. Phys.}, 43(12), 1975.

\bibitem{TFJordan69}
T.~F. Jordan.
\newblock {\em Linear Operators for Quantum Mechanics}.
\newblock John Wiley and Sons, Inc., 1969.

\bibitem{Ballentine98}
L.~E. Ballentine.
\newblock {\em Quantum Mechanics : A Modern Development}.
\newblock World Scientific, 1999.

\bibitem{Dickson-Dirac-talk}
M.~Dickson.
\newblock Beauty doth of itself persuade.
\newblock Presentation at University of South Carolina (overheads at
  {www.mdickson.com/pubs}), 2005.

\bibitem{Goyal-QT2c}
P.~Goyal.
\newblock Information-geometric reconstruction of quantum theory.
\newblock {\em Phys. Rev. A}, 78(5):052120, 2008.

\bibitem{Frieden-Schroedinger-derivation}
B.~R. Frieden and B.~H. Soffer.
\newblock Lagrangians of physics and the game of {F}isher-information transfer.
\newblock {\em Phys. Rev.~{E}}, 52:2274--2286, 1995.

\bibitem{MJWHall-Schroedinger-derivation}
M.J.W. Hall and M.~Reginatto.
\newblock Schroedinger equation from an exact uncertainty principle.
\newblock {\em J. Phys.~{A}}, 35(14):3289--3303, 2002.

\bibitem{Reginatto-Schroedinger-derivation}
M.~Reginatto.
\newblock Derivation of the equations of nonrelativistic mechanics using the
  principle of minimum {F}isher information.
\newblock {\em Phys. Rev.~{A}}, 58:1775--8, 1998.

\bibitem{Bohr-hydrogen-atom}
N.~Bohr.
\newblock On the constitution of atoms and molecules, part {I}.
\newblock {\em Phil. Mag.}, 26(1--24), 1913.

\bibitem{Bohr-line-spectra}
N.~Bohr.
\newblock On the quantum theory of line spectra.
\newblock {\em Z. Phys.}, 2(5):423--478, 1920.

\bibitem{sources-of-quantum-mechanics-introduction}
B.~L. van~der Waerden.
\newblock Editor's introduction.
\newblock In B.~L. van~der Waerden, editor, {\em Sources of Quantum Mechanics}.
  Dover Publications, 1967.

\bibitem{Ehrenfest-average-values}
P.~Ehrenfest.
\newblock Bemerkung {\"{u}}ber die angen{\"{a}}herte g{\"{u}}ltigkeit der
  klassischen mechanik innerhalb der quantenmechanik.
\newblock {\em Z. Phys.}, 45(7--8):455--457, 1927.

\bibitem{vNeumann55}
J.~von Neumann.
\newblock {\em Mathematical Foundations of Quantum Mechanics}.
\newblock Princeton {U}niversity {P}ress, 1955.

\bibitem{Groenewold-Principles-QM}
H.~J. Groenewold.
\newblock On the principles of elementary quantum mechanics.
\newblock {\em Physica}, 12:405--60, 1946.

\bibitem{Bohm51}
David Bohm.
\newblock {\em Quantum Theory}.
\newblock Dover Publications, 1989.

\bibitem{Isham-QT}
C.~J. Isham.
\newblock {\em Lectures on Quantum Theory}.
\newblock Imperial College Press, 1995.

\bibitem{Isham-quantum-and-reals}
C.~J. Isham.
\newblock Some reflections on the status of conventional quantum theory when
  applied to quantum gravity.
\newblock Jun 2002.

\bibitem{Group-Theory-in-Physics}
J.~F. Cornwell.
\newblock {\em Group Theory in Physics: An Introduction}.
\newblock Academic Press, 1997.

\end{thebibliography}
\end{document}